\documentclass[reprint,
 superscriptaddress,
 amsmath,
 amssymb,
 aps,
 prx,
 floatfix,
 twocolumn
]{revtex4-1}

\newcommand{\bra}[1]{\langle #1\rvert}
\newcommand{\ket}[1]{\lvert #1\rangle}
\newcommand{\ip}[2]{\bra{#1} #2\rangle}
\newcommand{\op}[2]{\ket{#1} \bra{#2}}
\newcommand{\exv}[1]{\langle #1\rangle}

\newcommand{\pd}[1]{\frac{\partial #1}{\partial t}}
\newcommand{\rpd}[1]{\partial_t #1}

\DeclareMathOperator{\Tr}{Tr}
\DeclareMathOperator{\diag}{diag}

\usepackage{graphicx}% Include figure files
\usepackage{dcolumn}% Align table columns on decimal point
\usepackage{bm}% bold math
\usepackage{hyperref}% add hypertext capabilities
\usepackage{comment}% add comment environment
\usepackage{tikz}
\usetikzlibrary{calc}

\usepackage{amsmath}
\usepackage{upgreek}
\usepackage{physics}
\usepackage{isomath}
\usepackage{placeins}

\begin{document}

%\preprint{APS/123-QED}

\title{Adiabatic Control of Decoherence-Free-Subspaces in an Open Collective System}

\author{Jarrod T. Reilly}
\thanks{Corresponding author: Jarrod.Reilly@colorado.edu}
\affiliation{JILA, NIST, and Department of Physics, University of Colorado, 440 UCB, Boulder, CO 80309, USA}
\author{Simon B. J\"ager}
\affiliation{JILA, NIST, and Department of Physics, University of Colorado, 440 UCB, Boulder, CO 80309, USA}
\affiliation{Physics Department and Research Center OPTIMAS, Technische Universit\"at Kaiserslautern, D-67663, Kaiserslautern, Germany}
\author{John Cooper}
\affiliation{JILA, NIST, and Department of Physics, University of Colorado, 440 UCB, Boulder, CO 80309, USA}
\author{Murray J. Holland}
\affiliation{JILA, NIST, and Department of Physics, University of Colorado, 440 UCB, Boulder, CO 80309, USA}

\date{\today}

\pacs{Valid PACS appear here}% PACS, the Physics and Astronomy
\begin{abstract}
We propose a method to adiabatically control an atomic ensemble using a decoherence-free subspace (DFS) within a dissipative cavity.
We can engineer a specific eigenstate of the system's Lindblad jump operators by injecting a field into the cavity which deconstructively interferes with the emission amplitude of the ensemble.
In contrast to previous adiabatic DFS proposals, our scheme creates a DFS in the presence of \emph{collective} decoherence.
We therefore have the ability to engineer states that have high multi-particle entanglements which may be exploited for quantum information science or metrology.
We further demonstrate a more optimized driving scheme that utilizes the knowledge of possible diabatic evolution gained from the so-called adiabatic criteria.
This allows us to evolve to a desired state with exceptionally high fidelity on a time scale that does not depend on the number of atoms in the ensemble.
By engineering the DFS eigenstate adiabatically, our method allows for faster state preparation than previous schemes that rely on damping into a desired state solely using dissipation.
\end{abstract}

{
\let\clearpage\relax
\maketitle
}

\section{Introduction}
The control of quantum systems is at the core of many groundbreaking scientific advancements.
For example, laser cooling and trapping~\cite{Metcalf} has given rise to optical tweezers~\cite{Ashkin} and realizations of atomic condensates~\cite{Anderson,Regal}.
Other pioneering fields such as quantum computation and metrology rely on quantum control to prepare and study useful quantum states which can be seen as quantum state engineering.
This has led to rapid progress in quantum supremacy experiments~\cite{Schulte,Arute} and tests of relativity using atomic clocks~\cite{Chou,Hutson,Bothwell}.
A common procedure to achieve the desired state evolution in these quantum platforms with high fidelity is to require that the dynamics during the state engineering process remain adiabatic~\cite{Aharonov,Albash,Pang,Venegas-Gomez,Jin}.

However, the investigation of such controlled quantum systems as platforms for quantum computation, memory, metrology, and simulation is often limited experimentally due to decoherence induced by the system's coupling to its environment~\cite{Joos,Gardiner,Orszag,DiVincenzo,Unrah,Shor,Suter,Roos,Ostermann,Sekatski,Dur,Zuniga-Hansen,Hauke}.
As a result, many schemes have been developed to evolve the system in a dark state so that the system does not undergo any non-unitary evolution~\cite{Steck}.
Common procedures for achieving this in the presence of spontaneous emission are stimulated Raman transitions~\cite{Kasevich,Kasevich2,Moler,Reilly}, stimulated Raman adiabatic passage (STIRAP)~\cite{Oreg,Weitz,Vitanov} and electromagnetically induced transparency (EIT)~\cite{Tewari,Harris,Boller} which all are coherent schemes that utilize quantum interference effects to achieve dynamics entirely in the ground state manifold.
However, procedures that rely entirely on coherent dynamics restrict the scope of what may be studied. This is because superpositions between electronic ground and excited states cannot be maintained due to the decay of coherences from the finite lifetime of the excited state.
Furthermore, these schemes are single-particle control procedures and therefore cannot readily be used to generate, in a controllable manner, collective states with multi-particle correlations that might be exploited.
To overcome this limitation, one can instead prepare the system in a dissipative dark state in which deconstructive interference allows a system to evolve with a suppressed rate of decoherence even in the presence of a large excited state population.
An example of this are many-body subradiant states~\cite{Dicke,Guerin,Weiss,Ostermann,Shankar}, although these states are often not pure.

A well-studied protocol to create dark states that remain pure in the presence of decoherence is the preparation of the system in a so-called decoherence-free subspace (DFS).
Here, a system remains pure because it is constructed to be in a subspace spanned by the eigenstates of all of the system's Lindblad jump operators and therefore undergoes solely coherent dynamics (i.e. noiseless evolution) within this subspace~\cite{Lidar,Zanardi,Lidar2}.
Counterintuitively, decoherence in the presence of a DFS \emph{generates} coherences between the DFS eigenstates and states outside the DFS manifold so that the system tends to damp into the DFS, as exploited in~\cite{DallaTorre}.
In addition, a Hamiltonian may be added to exactly cancel these coherences in order to create a dynamically stable DFS~\cite{Karasik}.
This makes the use of DFS a promising tool for realizations of quantum metrology and information procedures~\cite{Roos,DallaTorre,Ostermann,Sekatski,Hamann,Kielpinski,Brooke,Patra,He}.
However, the system may have a long relaxation time and therefore increase the chance of other sources of experimental noise to become relevant.

Combining the consideration of quantum control, stability to decoherence, and evolution time, it is thus desirable to create a DFS adiabatically, and procedures have been proposed for achieving this~\cite{Carollo,Wu,Wu2}.
In these procedures, a system will adiabatically follow the DFS eigenstates provided a so-called adiabaticity criteria is satisfied~\cite{Wu}.
To our knowledge, these adiabatic DFS schemes have only been applied for single-particle control and therefore have been limited in scope.
In this paper, we introduce a scheme to adiabatically control an atomic spin ensemble interacting with a highly dissipative cavity. 
By driving the cavity with a light field that deconstructively interferes with the emission amplitude of the atoms, we can engineer the system to evolve into a specific eigenstate of the system's jump operator.
Furthermore, we use the adiabatic criteria to develop a driving scheme that can achieve extremely high fidelities on short time scales even in the presence of large atom numbers.
As a specific example, we demonstrate how to create a metrologically useful dissipative state, that is similar to the one studied in~\cite{DallaTorre}, adiabatically which is advantageous as this state can not be obtained simply by a sudden parameter quench.

The article is organized as follows.
We begin with a review on the requirements for a dynamically stable DFS in Sec.~\ref{DyanmicallyStableDFS}.
We then derive the collective atomic-cavity interaction model in Sec.~\ref{SpinFlipModel}. In Sec.~\ref{DiagonalizationOfJumpOperator}, we introduce a method to compute the jump operator's eigenvectors that will define the DFS as well as its orthogonal complement. 
After that, in Sec.~\ref{AdiabaticDFS}, we show an adiabatic protocol to prepare the atom in the DFS. We conclude with an outlook and discussion of future work in Sec.~\ref{ConclusionOutlook}.

\section{Dynamically Stable Decoherence-Free-Subspaces} \label{DyanmicallyStableDFS}
We first briefly review the criteria for a pure and dynamically stable DFS eigenstate that we consider throughout the paper. The dynamics of the density operator $\hat{\rho}(t)$, describing the studied quantum states, is governed by a Born-Markov master equation
\begin{equation}
    \pd{\hat{\rho}} = \hat{\mathcal{L}} \hat{\rho} := \frac{1}{i \hbar} \left[ \hat{H}, \hat{\rho} \right] + \sum_i \hat{\mathcal{D}} \left[ \hat{L}_i \right] \hat{\rho},
\end{equation}
where $\hat{\mathcal{L}}$ is the Liouvillian superoperator.
The coherent dynamics is governed by the Hamiltonian $\hat{H}$ and the dissipation is described by the Lindblad jump operators $\hat{L}_i$ using the Lindblad superoperator
\begin{equation} \label{LindbladSuperoperator}
   \hat{\mathcal{D}} \left[ \hat{L}_i \right] \hat{\rho} = \hat{L}_i \hat{\rho} \hat{L}_i^{\dagger} - \frac{1}{2} \left( \hat{L}_i^{\dagger} \hat{L}_i \hat{\rho} + \hat{\rho} \hat{L}_i^{\dagger} \hat{L}_i \right).
\end{equation}

As formulated in~\cite{Karasik,Wu}, a dynamically stable DFS in which the basis states remain pure, e.g. $\hat{\rho}^2(t) = \hat{\rho} (t)$, is defined by two necessary and sufficient conditions.
\begin{enumerate}
\item The first condition is the general Lidar-Chuang-Whaley theorem~\cite{Lidar} which requires that all basis states $\{ \ket{m} \}$ of a DFS $\mathcal{H}_{\text{DFS}}=\text{span} \left[ \{\ket{m}\} \right]$ are degenerate eigenstates of all $\hat{L}_i$:
\begin{equation}
    \hat{L}_i \ket{m} = \Lambda_i \ket{m}, 
\end{equation}
for every $i$ and $\ket{m} \in \mathcal{H}_{\text{DFS}}$.

\item The second condition requires that $\mathcal{H}_{\text{DFS}}$ is invariant to the effective Hamiltonian
\begin{equation} \label{H_eff}
    \hat{H}_{\text{eff}} = \hat{H} + \frac{i \hbar}{2} \sum_i \left[ \Lambda_i^*  \hat{L}_i  - \Lambda_i  \hat{L}_i^{\dagger}\right],
\end{equation}
such that the DFS basis states satisfy the condition
\begin{equation} \label{DFScondition}
    \bra{n^{\perp}} \hat{H}_{\text{eff}} \ket{m} = 0,
\end{equation}
for every $\ket{m} \in \mathcal{H}_{\text{DFS}}$ and $\ket{n^{\perp}} \in \mathcal{H}_{\text{CS}}$ where $\mathcal{H}_{\text{CS}}$ is the orthogonal complement of $\mathcal{H}_{\text{DFS}}$.
\end{enumerate}

In this paper, we consider dynamically varying the Hamiltonian $\hat{H}=\hat{H}(t)$ and jump operators $\hat{L}_j=\hat{L}_j(t)$.
Consequently, the DFS will also change in time and one can ask whether we will dynamically stay in a DFS during the system's evolution. 
We now follow Ref.~\cite{Wu} where it has been shown that a pure state initialized in the DFS at $t=0$ will remain in the DFS provided that the basis of $\mathcal{H}_{\text{DFS}}$ and $\mathcal{H}_{\text{CS}}$ are continuous with time and fulfill the adiabatic condition
\begin{equation} \label{AdiabaticityParameterDef}
    \Xi (t) = \max_{m,n} \abs{\frac{4 \bra{n^{\perp}} \rpd \ket{m}}{\alpha_{mn} + i \zeta_n}} \ll 1,
\end{equation}
for every $\ket{m} \in \mathcal{H}_{\text{DFS}}$ and $\ket{n^{\perp}} \in \mathcal{H}_{\text{CS}}$. 
In Eq.~\ref{AdiabaticityParameterDef} we have introduced $\hbar \alpha_{mn} = \bra{n^{\perp}} \hat{H}_{\text{eff}} \ket{n^{\perp}} - \bra{m} \hat{H}_{\text{eff}} \ket{m}$ and $\zeta_n = \sum_i \bra{n^{\perp}} (\hat{L}_i^{\dagger} - \Lambda_i^*) (\hat{L}_i - \Lambda_i) \ket{n^{\perp}}/2$.
In the following sections, we will use these conditions to derive a unique driving profile in a collective spin system to adiabatically follow a many-body DFS eigenstate.

\section{Collective Spin-Flip Model} \label{SpinFlipModel}
\begin{figure}
    \centerline{\includegraphics[width=\linewidth]{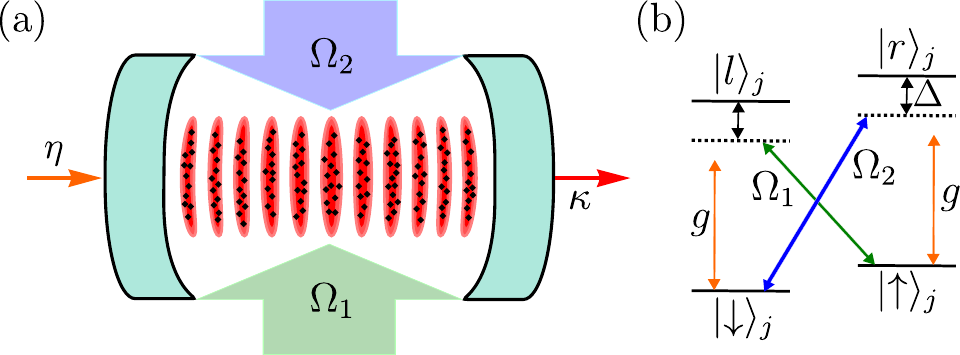}}
    \caption{(a) Schematic diagram of our system with $N$ atoms trapped at the anti-nodes of a cavity. (b) Level diagram of the four-level internal structure of atom $j$.}
    \label{Model}
\end{figure}
We now study the collective spin system shown in Fig.~\ref{Model}(a).
We consider $N$ identical four-level atoms that couple to a single mode of an optical cavity with identical coupling constant $g$. 
This can be achieved by trapping the atoms at the antinodes of the cavity mode function. 
Cavity photons with frequency $\omega_c$ decay into free space at rate $2 \kappa$ and are driven externally by a laser field with pump strength $\eta$ through the cavity mirrors.
The atoms are also driven by two additional laser fields $\Omega_1$ and $\Omega_2$, with frequencies $\omega_1$ and $\omega_2$ respectively, that couple different states than the cavity field [see Fig.~\ref{Model}(b)]. 
The internal structure of atom $j$ is depicted schematically in Fig.~\ref{Model}(b), where each atom has two ground states $\ket{\downarrow}$ and $\ket{\uparrow}$, and two excited states $\ket{l}$ and $\ket{r}$ with bare frequencies $\omega_l$ and $\omega_r$ with respect to the frequency of $\ket{\downarrow}$.
Note that the correct couplings with the cavity and classical fields can be accomplished in physical systems using hyperfine split states~\cite{Dimer,Masson}, states in different hyperfine manifolds~\cite{DallaTorre,Grimsmo,Zhang,Masson2}, or with two-component Bose-Einstein condensates~\cite{Kroeze,Mivehvar}.

In the regime where the driving lasers corresponding to $\Omega_1$ and $\Omega_2$ are off-resonant, $\Delta_l, \Delta_r \gg \Omega_1, \Omega_2$, we eliminate the two excited states $\ket{l}$ and $\ket{r}$ resulting in an effective master equation for $N$ two level atoms with states $\ket{\downarrow}$ and $\ket{\uparrow}$, that couple to the single cavity mode.
Here, we defined the large detunings $\Delta_l = (\omega_1 + \omega_2)/2 - \omega_{l}$ and $\Delta_r = \omega_2 - \omega_{r}$ between the driving lasers and the upper-state manifold. 
Thereafter, we eliminate the cavity mode assuming that the typical lifetime of a cavity photon is much shorter than the typical timescale of a collectively enhanced two-photon Raman process, $\kappa \gg \sqrt{N} \eta, \sqrt{N} S_1, \sqrt{N} S_2$, with $S_1 = g \abs{\Omega_1}/(2 \abs{\Delta_l})$ and $S_2 = g \abs{\Omega_2}/(2 \abs{\Delta_r})$. 
The result of this calculation is an effective master equation describing the driven-dissipative dynamics of $N$ two-level atoms with states $\ket{\downarrow}$ and $\ket{\uparrow}$.
For details of the derivation of the effective master equation, we refer the reader to Appendix~\ref{MasterEquationDerivation}.

This effective master equation governs the dynamics of the atomic density matrix $\hat{\rho}_{\text{at}}$ and reads
\begin{equation} \label{EffectiveMasterEq}
    \pd{\hat{\rho}_{\text{at}}} = \hat{\mathcal{L}}_{\text{at}} \hat{\rho}_{\text{at}} := \frac{1}{i\hbar} \left[ \hat{H}_{\text{at}}, \hat{\rho}_{\text{at}} \right] + \hat{\mathcal{D}} \left[ \hat{L} \right] \hat{\rho}_{\text{at}},
\end{equation}
with the effective jump operator
\begin{equation} \label{FullJumpOp}
    \hat{L} = \sqrt{\Gamma_c} \left( \hat{J}^- + \mu^2 \hat{J}^+ + \chi \hat{\mathbb{I}} \right),
\end{equation}
and the effective Hamiltonian given by
\begin{equation} \label{H_at}
    \hat{H}_{\text{at}} = \frac{\hbar \nu}{2} \hat{L}^{\dagger} \hat{L}.
\end{equation}
Here, we have used the definition of the collective raising $\hat{J}^+$ and lowering oprators $\hat{J}^-$ defined by
\begin{equation}
    \hat{J}^+ = \sum_{j=1}^N \ket{\uparrow}_j \bra{\downarrow}_j = \left( \hat{J}^- \right)^{\dagger}.
\end{equation}
The rate 
\begin{equation} \label{GammaC}
    \Gamma_c = \frac{2 \kappa S_1^2}{\Delta_c^2+\kappa^2}
\end{equation}
is the cavity-induced spontaneous emission rate from $\ket{\uparrow} $ to $\ket{\downarrow}$, with the cavity detuning $\Delta_c=(\omega_1+\omega_2)/2-\omega_{c}$. In addition we defined the ratios
\begin{equation} \label{MuAndChi}
    \mu = \sqrt{\frac{S_2}{S_1}}, \quad \chi = \frac{\eta}{S_1},
\end{equation}
and $\nu = \Delta_c/\kappa$, where we have made an assumption of the phase of $\eta$ as discussed in Appendix~\ref{RotatingFrame}.

\begin{figure}
    \centerline{\includegraphics[width=0.7\linewidth]{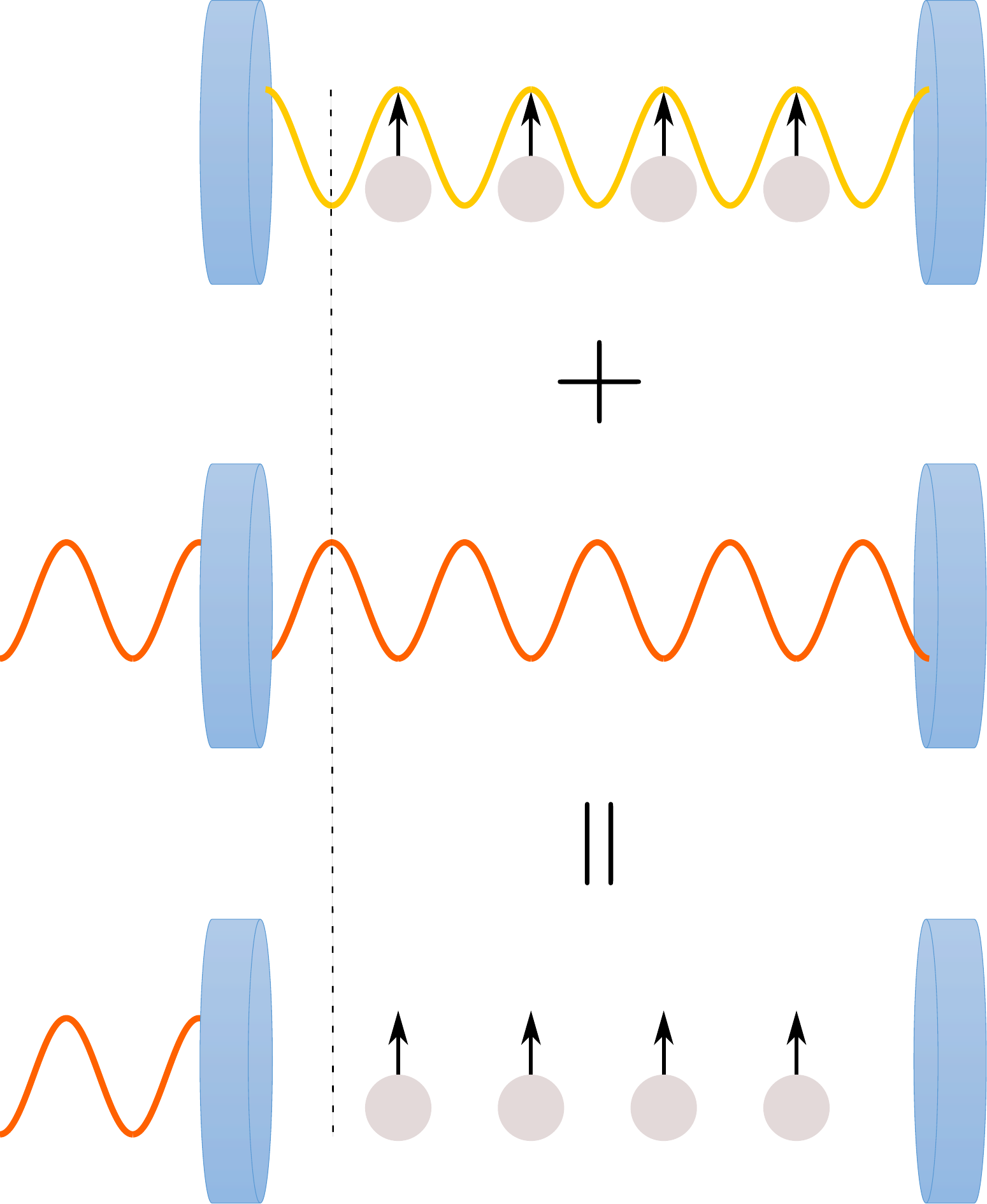}}
    \caption{Sketch of the general idea behind engineering a dark state in a cavity.
    The emission amplitude of the atomic state (top) is canceled by an external driving laser (middle) resulting in a zero photon field (bottom).
    Note that this effectively makes the atom-cavity system a perfectly reflective mirror for the external driving light.}
    \label{DarkCavity}
\end{figure}
We can now apply the results that we have reviewed in Sec.~\ref{DyanmicallyStableDFS}. 
In fact, it is rather easy to see that both conditions to obtain a DFS are fulfilled by a dark state $\ket{\Psi_{\mathrm{D}}}$ of $\hat{L}$ with
\begin{equation} \label{DarkState}
    \hat{L} \ket{\Psi_{\mathrm{D}}} = 0,
\end{equation}
because we directly obtain $\hat{\mathcal{L}}_{\text{at}} \op{\Psi_{\mathrm{D}}}{\Psi_{\mathrm{D}}} = 0$. 
As a direct result, all states that fulfill Eq.~\eqref{DarkState} span a DFS and we can engineer this state by modifying the ratios $\mu$ and $\chi$ of the external driving lasers. 
This rather mathematical description of the system has a very simple physical explanation. 
An atomic ensemble in an eigenstate $\ket{\Psi}$ of $\hat{L}$ gives rise to a cavity field with a certain amplitude $a$. 
This amplitude depends on the atomic state and is determined by $a \propto \bra{\Psi} \tilde{L} \ket{\Psi}$ with
$\tilde{L} = \hat{J}^- + \mu^2 \hat{J}^+$.
By shining in a light field $a_{\mathrm{ext}}\propto \chi$ that destructively interferes with the cavity field, such that $a+a_{\mathrm{ext}}=0$, the atomic state remains in a dark state and is therefore unperturbed by the cavity field [see Fig.~\ref{DarkCavity}].

This perturbation only vanishes exactly if the atomic state is an eigenstate of $\hat{L}$ and therefore of $\tilde{L}$. 
The diagonalization of this operator is the topic of the next section. 
At this point, we want to remark that $\hat{L}$ commutes with the total length of the Bloch vector $\hat{J}^2$~\cite{Meiser}.
Since a natural initial state for this system is the state where all atoms are either in $\ket{\uparrow}$ or in $\ket{\downarrow}$, we restrict ourself to the state space within the manifold of $N+1$ symmetric Dicke states $\ket{J = N/2, m}$, with $m = -N/2, -N/2+1, \dots, N/2$.

\section{Diagonalization of the Jump Operator} \label{DiagonalizationOfJumpOperator}

\subsection{Schwinger Boson Representation}
For analytic ease in finding the dark state of Eq.~\eqref{FullJumpOp}, we utilize the Schwinger boson representation~\cite{Schwinger} to represent the symmetric Dicke states of the system.
Here, we introduce two modes with creation (annihilation) operators $\hat{b}_{\uparrow}^{\dagger}$ ($\hat{b}_{\uparrow}$) and $\hat{b}_{\downarrow}^{\dagger}$ ($\hat{b}_{\downarrow}$) which represent the ``creation'' (``annihilation'') of a particle in the states $\ket{\downarrow}$ and $\ket{\uparrow}$, respectively.
With this Schwinger boson representation, we can then write down the operator $\hat{L}$ as a non-Hermitian quadratic operator
\begin{equation}
    \hat{L} = \hat{\bf b} ^{\dagger} {\bf L} \hat{\bf b}+\chi,
\end{equation}
with 
\begin{equation}
    {\bf L} = 
    \begin{pmatrix}
    0 & \mu^2 \\ 
    1 & 0
    \end{pmatrix},
\end{equation}
and $\hat{\bf b} = (\hat{b}_{\uparrow}, \hat{b}_{\downarrow})^T$.
Although ${\bf L}$ is not Hermitian we can still diagonalize it if $\mu \neq 0$. 
In that case, we find the unnormalized eigenvectors ${\bf V}$ such that ${\bf L}=\boldsymbol{V} \boldsymbol{D} \boldsymbol{V}^{-1}$ with $\boldsymbol{D} = \diag(\mu, -\mu)$ being the eigenvalues of ${\bf L}$. 
The matrix of ${\bf V}$ is given by
\begin{equation}
    {\bf V}= \begin{pmatrix}
    \mu &-\mu \\
    1 &1
    \end{pmatrix}.
\end{equation}

\subsection{Generalized Eigenvectors} \label{EigenvectorSection}
Using the form of ${\bf V}$, we can then define the operators $\hat{\bf c} = {\bf V}^{\dagger} \hat{\bf b}$ and $\hat{\bf d} = {\bf V}^{-1} \hat{\bf b}$ with $\hat{\bf c} = (\hat{c}_{1}, \hat{c}_{2})^T$ and $\hat{\bf d} = (\hat{d}_{1}, \hat{d}_{2})^T$ such that we can rewrite 
\begin{equation}
    \hat{L}=\mu[\hat{c}_1^{\dag}\hat{d}_1-\hat{c}_2^{\dag}\hat{d}_2]+\chi.
\end{equation}
Using $[\hat{d}_i, \hat{c}_j^{\dag}] = \delta_{ij}$ we find the eigenvectors of the jump operator in Eq.~\eqref{FullJumpOp} in the general form
\begin{equation} \label{DFSEigenstates}
    \ket{\psi_k} = \mathcal{N}_k \left( \hat{c}_1^{\dagger} \right)^{N-k} \left( \hat{c}_2^{\dagger} \right)^k \ket{0},
\end{equation}
where $\mathcal{N}_k$ is a normalization factor that is derived in Appendix~\ref{OverlapCalculation} [see Eq.~\eqref{N_k}] and $k \in \{ 0,1,2, \ldots, N \}$.
That means there are $N+1$ eigenvectors that are, in general, non-orthogonal because $[\hat{c}_1,\hat{c}_2^{\dag}]\neq0$.
The eigenvectors have the eigenvalues
\begin{equation}
    \hat{L}\ket{\psi_k}=[\mu(N-2k)+\chi]\ket{\psi_k}.
\end{equation}
For a given $k$, we can now create a unique DFS that contains only one eigenvector $\mathcal{H}_{\text{DFS}, k} = \text{span} \left[ \{\ket{\psi_k}\} \right]$ by choosing $\chi=-\mu(N-2k).$ This consideration is only true for $\mu \neq 0$, while for $\mu=0$ there is only a single one-dimensional DFS corresponding to all atoms in the $\ket{\downarrow}$ state for the choice $\chi=0$.

\begin{figure}
    \centerline{\includegraphics[width=\linewidth]{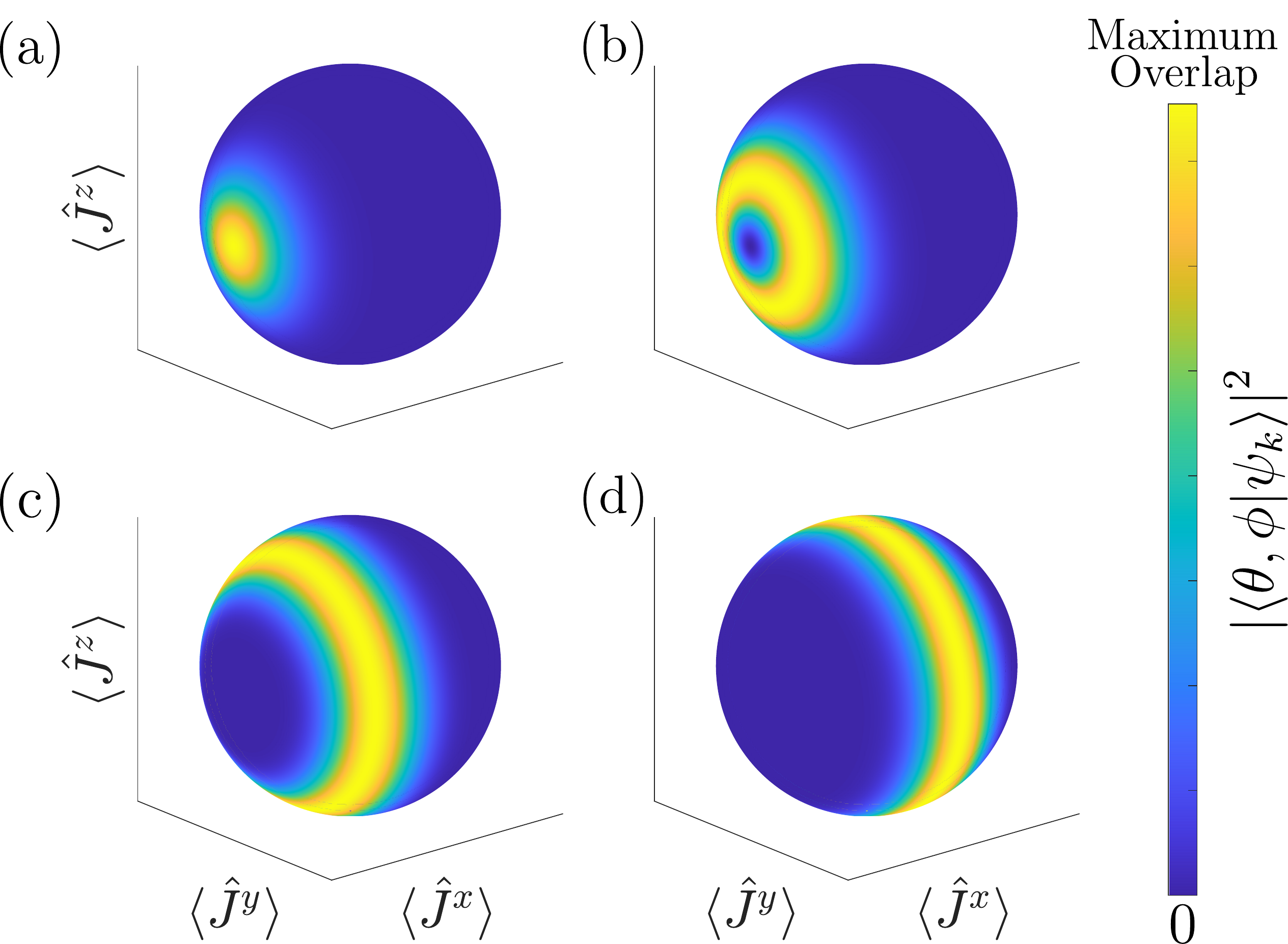}}
    \caption{The collective Bloch sphere of the DFS eigenstates $\ket{\psi_k}$ when $\mu = 1$. The color at each point is calculated by finding the overlap with the state at a certain point on the sphere, $\abs{\ip{\theta, \phi}{\psi_k}}^2$. We show the eigenstates (a) $k = 0$, (b) $k = 1$, (c) $k = N/4$, and (d) $k = N/2$ for an atom number $N = 20$. All distributions are normalized such that bright yellow regions represent states where the overlap is maximized while dark blue regions correspond to $\abs{\ip{\theta, \phi}{\psi_k}}^2 \approx 0$.}
    \label{CollectiveBlochSphere}
\end{figure}
To emphasize that the states in this DFS are in general non-trivial, coherent, and entangled states, we focus now on the case $\mu=1$. 
In that case, the eigenstates of the jump operator become eigenstates of $\hat{J}^x = \left( \hat{J}^+ + \hat{J}^- \right)/2$ which are useful for applications in quantum metrology. 
It is constructive to examine the different DFS eigenstates using the collective Bloch sphere, as shown in Fig.~\ref{CollectiveBlochSphere}.
Here, we calculate the overlap $\abs{\ip{\theta, \phi}{\psi_k}}^2$ of the $k^{\text{th}}$ DFS eigenstate with the spin coherent state,
\begin{equation}
    \ket{\theta, \phi} = \frac{1}{\sqrt{N!}} \left[ \cos \left( \frac{\theta}{2} \right) \hat{b}_{\uparrow} + \sin \left( \frac{\theta}{2} \right) e^{i \phi} \hat{b}_{\downarrow} \right]^N \ket{0},
\end{equation}
on the sphere's surface pointing in the direction given by its polar and azimuthal angles $\theta$ and $\phi$.
The $k = 0$ state, shown in Fig.~\ref{CollectiveBlochSphere}(a), represents a coherent spin state in which every atom is in the state $\ket{+}_j = \left( \ket{\uparrow}_j + \ket{\downarrow}_j \right)/\sqrt{2}$.
The $k = N$ state is the opposite coherent spin state with every atom in the $\ket{-}_j = \left( \ket{\uparrow}_j - \ket{\downarrow}_j \right)/\sqrt{2}$ state.
Meanwhile, for the eigenstates in between these extreme $k$ values, the ensemble becomes an entangled state that is a superposition of every permutation of $N-k$ atoms in $\ket{+}_j$ and $k$ atoms in $\ket{-}_j$.
Representing the eigenvectors on the collective Bloch sphere, we find vertical rings of varying radius with $\exv{\hat{J}^x}$ as its symmetry axis, as demonstrated in Figs.~\ref{CollectiveBlochSphere}(b) and (c) for the cases $k = 1$ and $k = N/4$, respectively.
The largest radius ring is the one corresponding to $k = N/2$ state which lies along the line of longitude at $\exv{\hat{J}^x} = 0$, as shown in Fig.~\ref{CollectiveBlochSphere}(d).
It consists of an equal number of atoms in $\ket{+}_j$ and $\ket{-}_j$ and is therefore naturally a dark state of the system, with a $\hat{J}^x$ eigenvalue of $0$. 
It has been demonstrated~\cite{DallaTorre} that the $k = N/2$ state for $\mu = 1 - \varepsilon$ with a small parameter $\varepsilon$ can be metrologically useful for atomic clocks as its variance in $\hat{J}^y$ scales at the Heisenberg limit.

\subsection{Adiabatic evolution and the orthogonal complement of DFS Eigenstates}
In the next section, we are interested in guiding the system dynamically through a DFS. Therefore it will be important that we dynamically assure that for a given $k\in\{0,...,N\}$, we have
\begin{equation}
    \chi(t)=-\mu(t)(N-2k) .  
\end{equation}
In addition, it is important to quantify if the system leaves the DFS and enters the orthogonal complement. 
Since $\hat{L}$ is a quadratic operator, we can find the orthogonal complement of the DFS eigenstate $\ket{\psi_k}$ as $\mathcal{H}_{\text{CS}, k} = \text{span} \left[ \{ \ket{\psi_n^{\perp}}, \; n \neq k \} \right]$, where we have defined
\begin{equation} \label{ComplementaryEigenstates}
    \ket{\psi_n^{\perp}} =  \left( 2 \mu \right)^N \mathcal{N}_n^{\perp} \left( \hat{d}_1^{\dagger} \right)^{N-n} \left( \hat{d}_2^{\dagger} \right)^n \ket{0},
\end{equation}
with a normalization $\mathcal{N}_n^{\perp}$ that is given by Eq.~\eqref{N_n_perp}.
These states satisfy the relation
\begin{equation} \label{PerpOverlap}
    \ip{\psi_n^{\perp}}{\psi_k} = \left( 2 \mu \right)^N \mathcal{N}_k \mathcal{N}_n^{\perp} (N-k)! k! \delta_{k,n}.
\end{equation}

\section{Adiabatic Decoherence-Free-Subspace} \label{AdiabaticDFS}
We assume throughout this section that the system begins the process in the collective ground state, $\ket{\downarrow}^N$, at $\mu(t=0) = 0$. 
This is the unique DFS and steady-state for $\chi(t=0)=0$, meaning that at $t=0$ the lasers driving the $\ket{\downarrow}\to\ket{r}$ transition and the cavity mode are switched off, $\Omega_2 = \eta = 0$. 
On the other hand, the laser driving the  $\ket{\uparrow}\to\ket{l}$ transition is switched on and will not be dynamically changed, $\Omega_1(t) = \text{const}$.
This results in a time independent value of $\Gamma_c$ [see Eq.~\eqref{GammaC}].

\subsection{Linear Scheme}
\begin{figure}
    \centerline{\includegraphics[width=\linewidth]{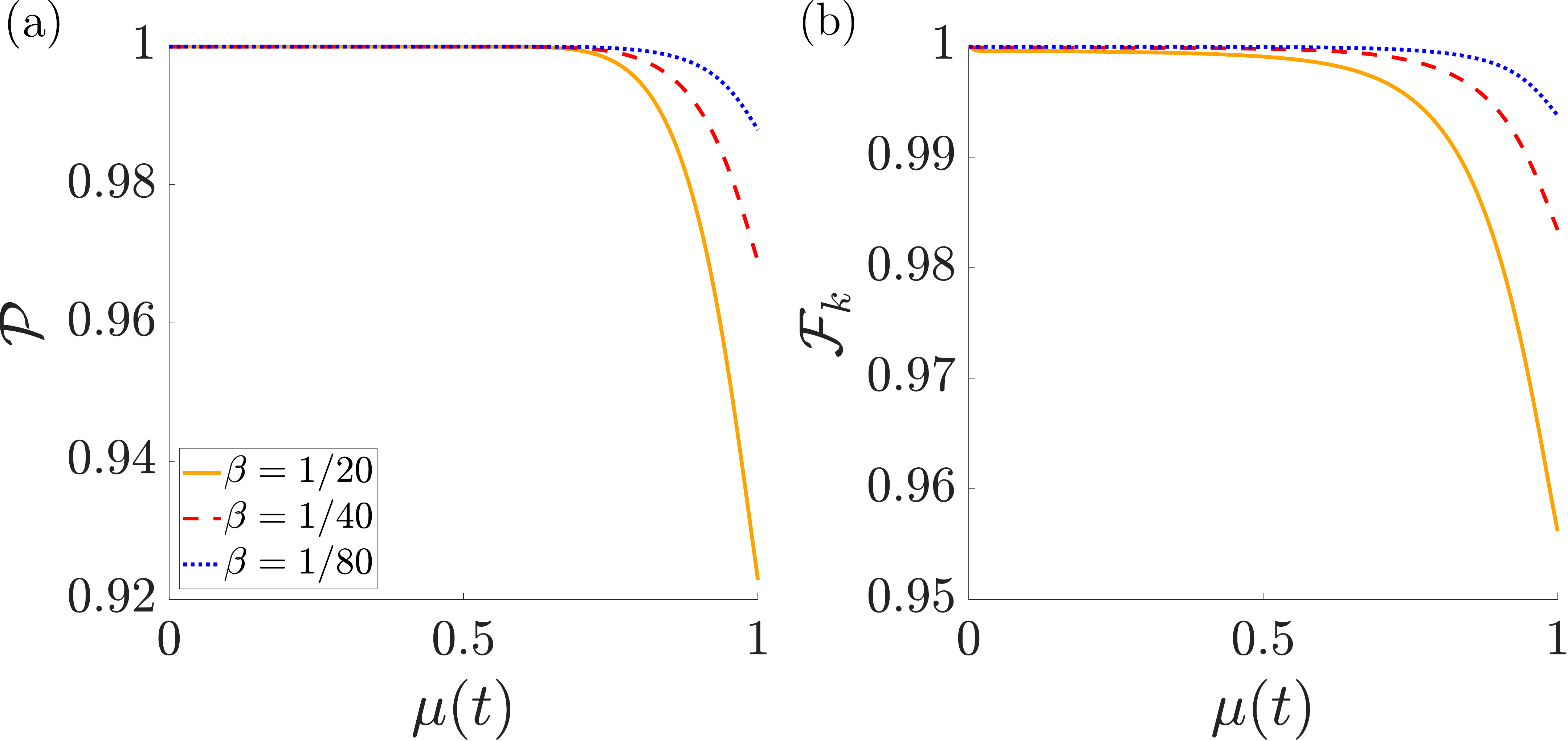}}
    \caption{(a) Purity $\mathcal{P}$ and (b) fidelity $\mathcal{F}_k$ during a linear sweep of $\mu$ for $\beta = 1/20$ (solid orange curve), $\beta = 1/40$ (dashed red curve), and $\beta = 1/80$ (dotted blue curve).
    Here, there are $N = 20$ atoms, we choose to create the state $k=0$, and set the parameters in units of $\Gamma_c$.
    We further assume $\nu = 0$.}
    \label{beta_loop}
\end{figure}
To exemplify the adiabatic creation of a desired DFS eigenstate, we first assume 
\begin{equation} \label{MuProfileLinear}
    \mu (t) = \beta t,
\end{equation}
with coefficient $\beta = 1/t_f$ such that $\abs{\Omega_1}$ is a constant and $\abs{\Omega_2}$ has a parabolic profile that reaches $\abs{\Omega_1} = \abs{\Omega_2}$ at the final time $t_f$.
The atomic state begins the process in its collective ground state $\hat{\rho}_{\text{at}} (0) = \op{N/2, -N/2}{N/2, -N/2}$ and we sweep $\mu$ with the desire that the atomic state finishes the process in a state that has a high overlap with the $k^{\text{th}}$ $\hat{J}^x$ eigenstate $\hat{\rho}_{\text{at}} (t_f) \approx \op{\psi_k (\mu = 1)}{\psi_k (\mu = 1)}$.
It stands to reason that the slower one sweeps $\mu$, the more adiabatic the dynamics become.
This intuition is demonstrated for three different values of $\beta$ in Fig.~\ref{beta_loop}. 
In Fig.~\ref{beta_loop}(a) we plot the purity $\mathcal{P} (t) = \Tr \left[ \hat{\rho}_{\text{at}}^2 \right]$ of the collective atomic state which becomes $\mathcal{P} = 1$ when the ensemble is in a pure state, while Fig.~\ref{beta_loop}(b) examines the Uhlmann-Jozsa fidelity~\cite{Jozsa}
\begin{equation} \label{Fidelity}
    \mathcal{F}_k = \left( \Tr \sqrt{\sqrt{\hat{\rho}_{\text{at}}} \op{\psi_k}{\psi_k} \sqrt{\hat{\rho}_{\text{at}}}} \right)^2 = \bra{\psi_k} \hat{\rho}_{\text{at}} \ket{\psi_k},
\end{equation}
of the dynamical atomic state with the desired instantaneous DFS eigenstate.
The plots illustrate the loss of both the final purity and final fidelity when $\beta$ is increased as diabatic dynamics causes the collective atomic state to dynamically transfer population from the desired $\ket{\psi_k}$ state to the neighboring eigenstates $\ket{\psi_{k \pm 1}}$.
Before studying this behavior in further detail, we first note that Fig.~\ref{beta_loop} suggests that a high final fidelity $\mathcal{F}_{k,f} \equiv \mathcal{F}_k (t_f)$ corresponds to a high final purity $\mathcal{P} (t_f)$ of the state.
However, a low fidelity does not necessarily correlate to a low purity as $\mu$ can be swept fast enough ($\beta \gg 1$) that the collective state remains approximately in its pure, but undesired, ground state $\hat{\rho}_{\text{at}} (t_f) \approx \op{N/2, -N/2}{N/2, -N/2}$.
We therefore choose to focus on the dynamical evolution of $\mathcal{F}_k$ to measure the level of success of our driving schemes for the rest of our analysis.

\begin{figure}
    \centerline{\includegraphics[width=\linewidth]{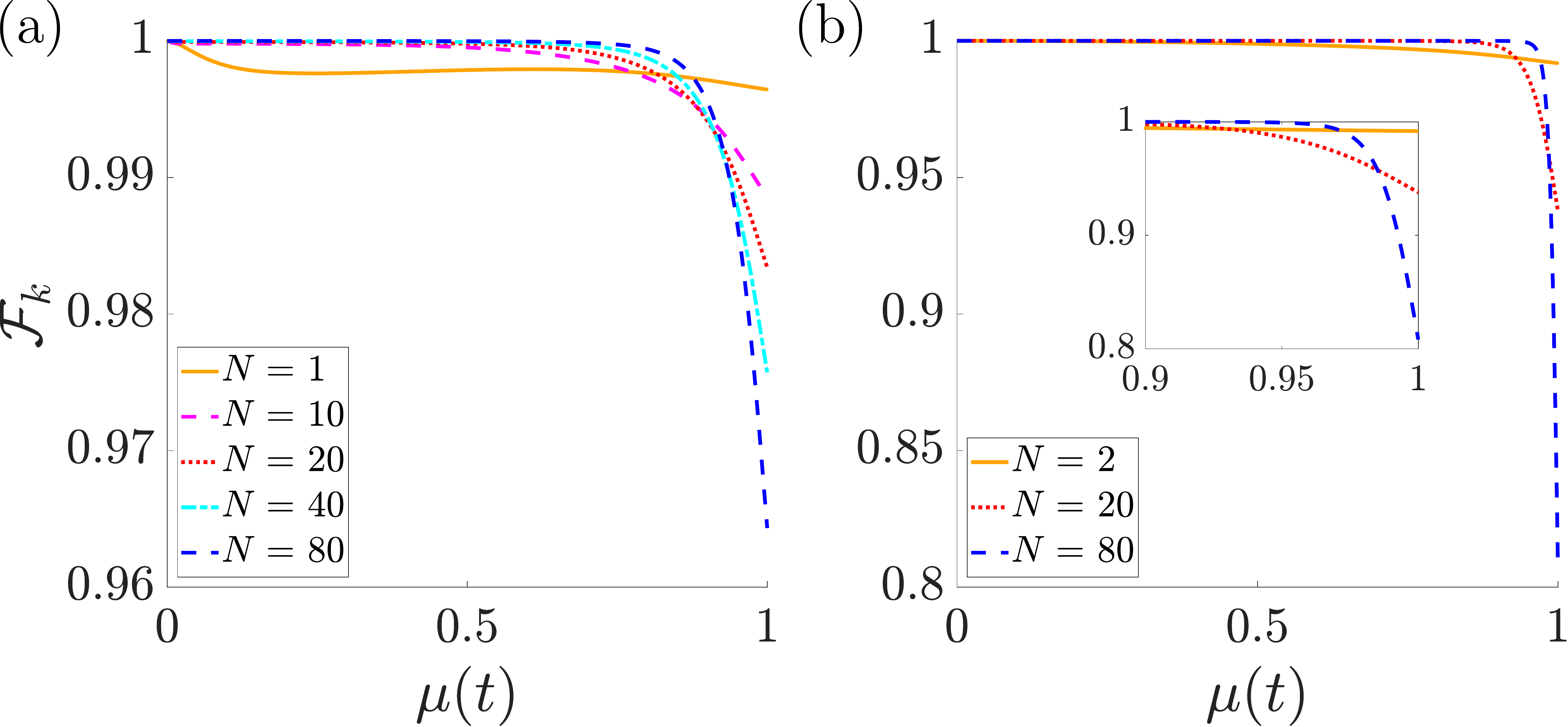}}
    \caption{Fidelity $\mathcal{F}_k$ with the states (a) $k=0$ and (b) $k = N/2$ for a linear scheme Eq.~\eqref{MuProfileLinear} with $t_f = 40/\Gamma_c$.
    We again have $\nu = 0$.
    The different curves represent different atoms numbers, with the small atom number $N = 1,2$ (solid orange curves), $N = 10$ (magenta dashed curve), $N = 20$ (dotted red curve), $N = 40$ (dotted-dashed cyan curve), and $N = 80$ (dashed blue curve). 
    The inset in (b) depicts the behavior of the fidelity for large values of $\mu$.}
    \label{N_loop_linear}
\end{figure}
An important question for experimental realizations of our adiabatic DFS scheme is how the loss of fidelity associated with non-adiabatic dynamics scales with the number of atoms in the ensemble $N$.
The results are displayed in Fig.~\ref{N_loop_linear} for (a) $k = 0$ and (b) $k = N/2$ when $t_f = 40/\Gamma_c$. We notice in both plots that as $N$ increases, the final fidelity $\mathcal{F}_{k,f}$ decreases rather significantly.
However, the dynamical evolution reveals that for increasing $N$, the state $\hat{\rho}_{\text{at}}$ remains approximately in the desired $\op{\psi_k}{\psi_k}$ state for a longer duration of the sweep before dropping to its lower final value.
The rate of the decay to $\mathcal{F}_{k,f}$ therefore becomes larger for increasing $N$.
To explain the behavior displayed in Fig.~\ref{N_loop_linear} in order to produce a driving scheme that can rectify the scaling of $\mathcal{F}_{k,f}$ with $N$, we now turn to the adiabaticity criteria introduced in Eq.~\eqref{AdiabaticityParameterDef}.

\subsection{Adiabatic Criteria}
The full calculation of the adiabaticity parameter is rather tedious and thus saved for Appendix~\ref{AdiabaticParameterCalculation} with the main results given by
\begin{equation} \label{Xi_k}
    \Xi_k = \frac{\dot{\mu}}{\Gamma_c\sqrt{1+\nu^2}} \xi_k,
\end{equation}
with the dimensionless parameter
\begin{equation} \label{xi_k}
    \xi_k = \max_{n = k \pm 1} \abs{\frac{4 \bra{\psi_{n}^{\perp}} \hat{J}^z \ket{\psi_k}}{\mu \left[ 2 \mu^2 + \left( 1 - \mu^4 \right) \bra{\psi_{n}^{\perp}} \hat{J}^z \ket{\psi_{n}^{\perp}} \right]}}.
\end{equation}
The maximization can be taken only over $n = k \pm 1$ since the explicit time derivative visible in Eq.~\eqref{AdiabaticityParameterDef} only couples to the ``neighbors'' of $k$ [see Eq.~\eqref{TimeDerivative}].
The quantity $\xi_k$ reaches its maximum for large values of $N$ close to $\mu=1$. 
We therefore define the value $\xi_k(\mu = 1)=\xi_{k,f}$ as it can be calculated analytically
\begin{equation} \label{xi_kf}
\xi_{k,f} = 
\begin{cases}
    & \sqrt{(N-k)(k+1)}, \quad k < \frac{N}{2} \\
    & \sqrt{(N-k+1)k}, \quad k \geq \frac{N}{2}
\end{cases}.
\end{equation}
For an adiabatic evolution we require $\Xi_k \ll 1$. 
The result of Eq.~\eqref{xi_kf} shows that $\xi_{k,f}$ increases with the number of atoms $N$.
Consequently, in order to fulfill the adiabatic criterion, $\Xi_k \ll 1$, we require $\dot{\mu}$ to be decreased with the atom number. 
This shows that a constant slope ramp, $\dot{\mu}= \text{const.}$, should fail in the large $N$ limit which is consistent with our findings in Fig.~\ref{N_loop_linear}.

\begin{figure}
    \centerline{\includegraphics[width=\linewidth]{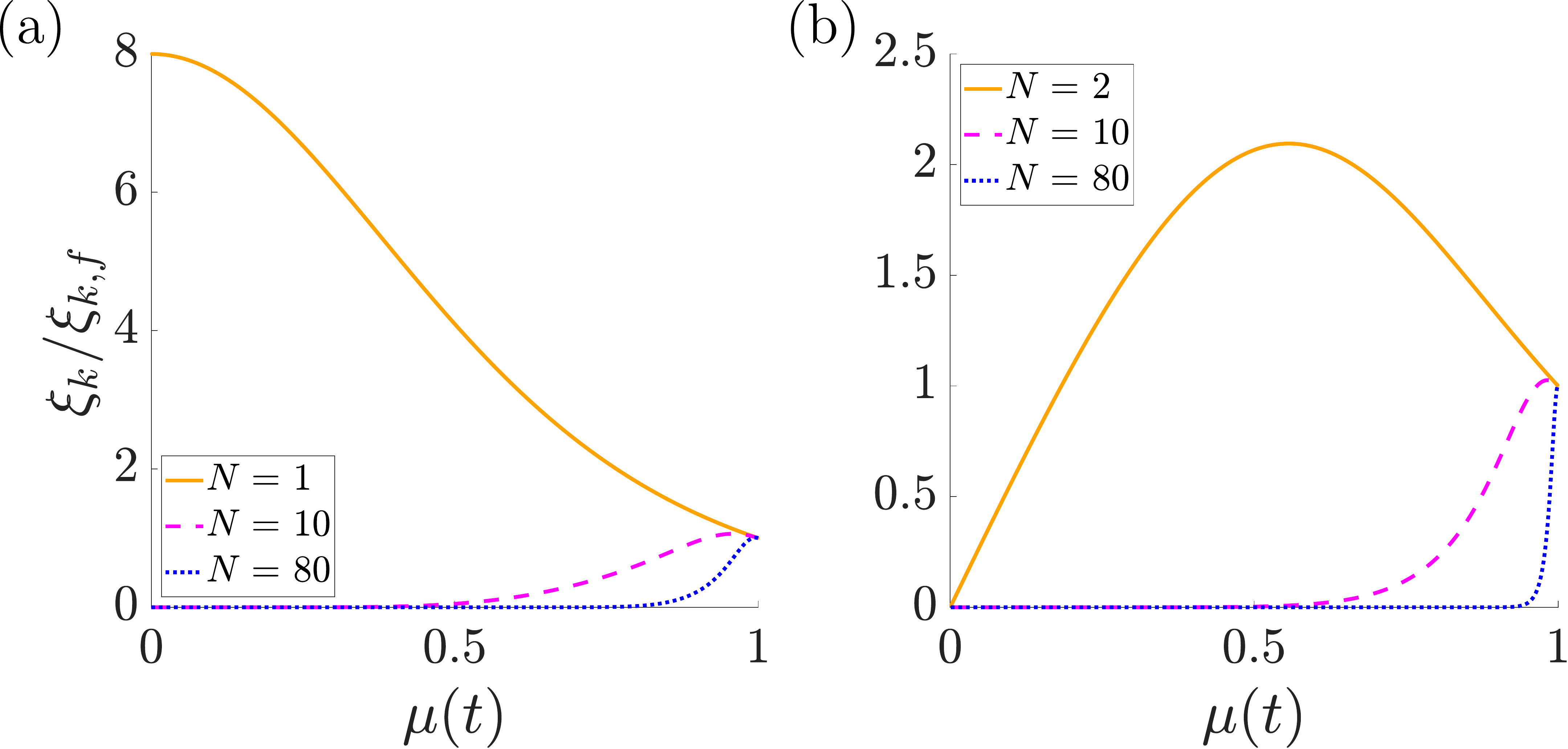}}
    \caption{The ratio $\xi_k/\xi_{k,f}$ [see Eqs.~\eqref{xi_k} and~\eqref{xi_kf}] as a function of $\mu$ for (a) $k = 0$ and (b) $k = N/2$.
    The curves are for small atom numbers $N = 1,2$ (solid orange curves), $N = 10$ (dashed magenta curve), and $N = 80$ (dotted blue curve).}
    \label{AdiabaticParameter_loop}
\end{figure}
To study this effect further, we now examine the adiabaticity parameter for values $\mu\in[0,1]$ in Fig.~\ref{AdiabaticParameter_loop} by plotting the ratio $\xi_k/\xi_{k,f}$ as a function of $\mu$ for different values of $N$ when $\nu = 0$.
Figure~\ref{AdiabaticParameter_loop}(a) shows the case $k = 0$.
For the example $N=1$, Eq.~\eqref{xi_kf} gives us $\xi_{k,f} = 1$ and the maximum value of the adiabaticity parameter clearly is obtained at $\mu = 0$ where we have $\xi_k = 8$.
Combining this analysis with the adiabaticity criteria $\Xi_k \ll 1$ and using the values of $t_f = 40/\Gamma_c$ that we have used in the previous subsection, we find that $\dot{\mu} = \Gamma_c/40 = \left[ 5 \max \left( \xi_k \right) \right]^{-1}$.
This choice of $\dot{\mu}$ is sufficient to satisfy the adiabaticity criteria $\dot{\mu} \ll \Gamma_c/\xi_k$ for the whole driving process such that adiabatic following of the desired state can occur.
Moreover, Figure~\ref{AdiabaticParameter_loop}(b) which displays the ratio for $k = N/2$ demonstrates that this choice of $\dot{\mu}$ is a suitable value for $N = 2$.
However, this value of $\dot{\mu}$ is no longer satisfactory for larger values of $N$ as the final value of $\xi_k$ scales as 
\begin{equation} \label{xi_kf_scaling}
    \begin{aligned} 
\xi_{k,f} \sim 
\begin{cases} \sqrt{N}, \quad k = 0 \\ 
\frac{N}{2}, \quad k = \frac{N}{2}
\end{cases},
    \end{aligned}
\end{equation}
leading to non-adiabticity which reduces $\mathcal{F}_{k,f}$ as $N$ increases which explains the behavior that was seen in Figure~\ref{N_loop_linear}.
This scaling can be interpreted as follows.
The time derivative of the eigenstates act as a raising and lowering operator to the two nearest eigenstates $k \pm 1$. 
It is this coupling that causes the diabatic evolution and since this coupling is induced through a cavity mode, it is collectively enhanced by the number of atoms in the cavity.
This coupling also depends on $k$ as it is larger for the eigenstates corresponding to $k \sim N/2$ than it is for the eigenstates on the edge $k \sim 0$ and $k \sim N$.
This can be explained in the full $2^N$ basis by noting that the number of permutations is $\binom{N}{k}$ so that middle states have many more individual atomic state combinations than the states on the ``edge''.
There are thus, in a sense, more avenues for the $k = N/2$ state to leak to the $k = N/2 \pm 1$ states than there are for the $k = 0$ state to leak to $k = 1$.
Note that $\xi_{k,f}$ is approximately the maximum value of $\xi_k$, which occurs slightly before $t_f$, and the relative difference between the maximum value and $\xi_{k,f}$ decreases with increasing $N$.

Another interesting feature displayed in Fig.~\ref{N_loop_linear} can now be explained using the adiabaticity criteria, $\Xi_k \ll 1$.
We demonstrated that the collective atomic state remains at a near perfect fidelity $\hat{\rho}_{\text{at}} \approx \op{\psi_k}{\psi_k}$ for a longer duration of the sweep of $\mu$ for increasing atom number. 
Figure~\ref{AdiabaticParameter_loop} clarifies this behavior as $\xi_k$ remains approximately zero, such that even choosing a very fast ramp, $\dot{\mu} \gg 1$, still can satisfy the adiabaticity criteria.
This fast ramp can be performed over a larger parameter space in $\mu$ if the number of atoms $N$ is increased. 
Close to $\mu \lesssim 1$, however, the value of $\xi_k$ rapidly increases to a value that grows with $N$ [see Eq.~\eqref{xi_kf_scaling}]. 
Thus, for this final stage, we have to choose a ramping speed $\dot{\mu}$ that is reduced with $N$ in order to remain adiabatic.
Somewhat counter-intuitively, the fastest dynamics can occur when the splitting $2 \mu$ between neighboring eigenstates in the jump operator's eigenspectrum is at its smallest while it must be slow when the splitting is largest.
This is because the overlap between neighboring eigenstates [given in Eq.~\eqref{EigenstateOverlaps}] is very large $\abs{\ip{\psi_{k \pm 1}}{\psi_k}}^2 \sim 1$ for a majority of the evolution before rapidly decreasing towards its final value of zero.
All these observations now allow us to construct a simple yet efficient driving scheme that produces a value of $\mathcal{F}_{k,f}$ that is robust as the atom number increases.

\subsection{Quench Scheme}
The analysis of the adiabatic parameter in the previous subsection implies that the dynamical evolution of $\mu$ can be very rapid for small values of $\mu$ before reaching a point where $\xi_k$ becomes very large.
This suggests that a constant $\dot{\mu}$ profile is not the most efficient driving profile.
Instead, it is sufficient to simply use a continuous piecewise linear $\mu$ profile which has an extremely steep initial slope when $\xi_k \ll 1$ and then becoming very gradual around the time when $\xi_k$ suddenly increases towards $\xi_{k,f}$. 
In the extreme limit, we have a scheme which quenches to a value of $\mu$ at $t = 0$ and then gradually evolving the system for the rest of the process until $\mu = 1$:
\begin{equation} \label{MuProfileQuench}
    \mu = 
    \begin{cases}
        & 0, \quad t < 0, \\
        & \beta_q t + C, \quad 0 \leq t \leq t_f,
    \end{cases}
\end{equation}
where $C = 1 - \beta_q t_f$.
With the choice $\beta_q = q/(t_f N)$, we can use Eq.~\eqref{xi_kf} to choose a $t_f$ and constant $q$ that satisfies $\beta_q \ll 1/\xi_k$ such that the dynamics remain adiabatic during the sweep.
Therefore, the only significant diabatic evolution occurs at the $t = 0$ when $\mu$ jumps to 
\begin{equation} 
    C = 1 - q/N,
\end{equation} 
which is quickly rectified by the system damping back into the DFS eigenstate.

\begin{figure}
    \centerline{\includegraphics[width=\linewidth]{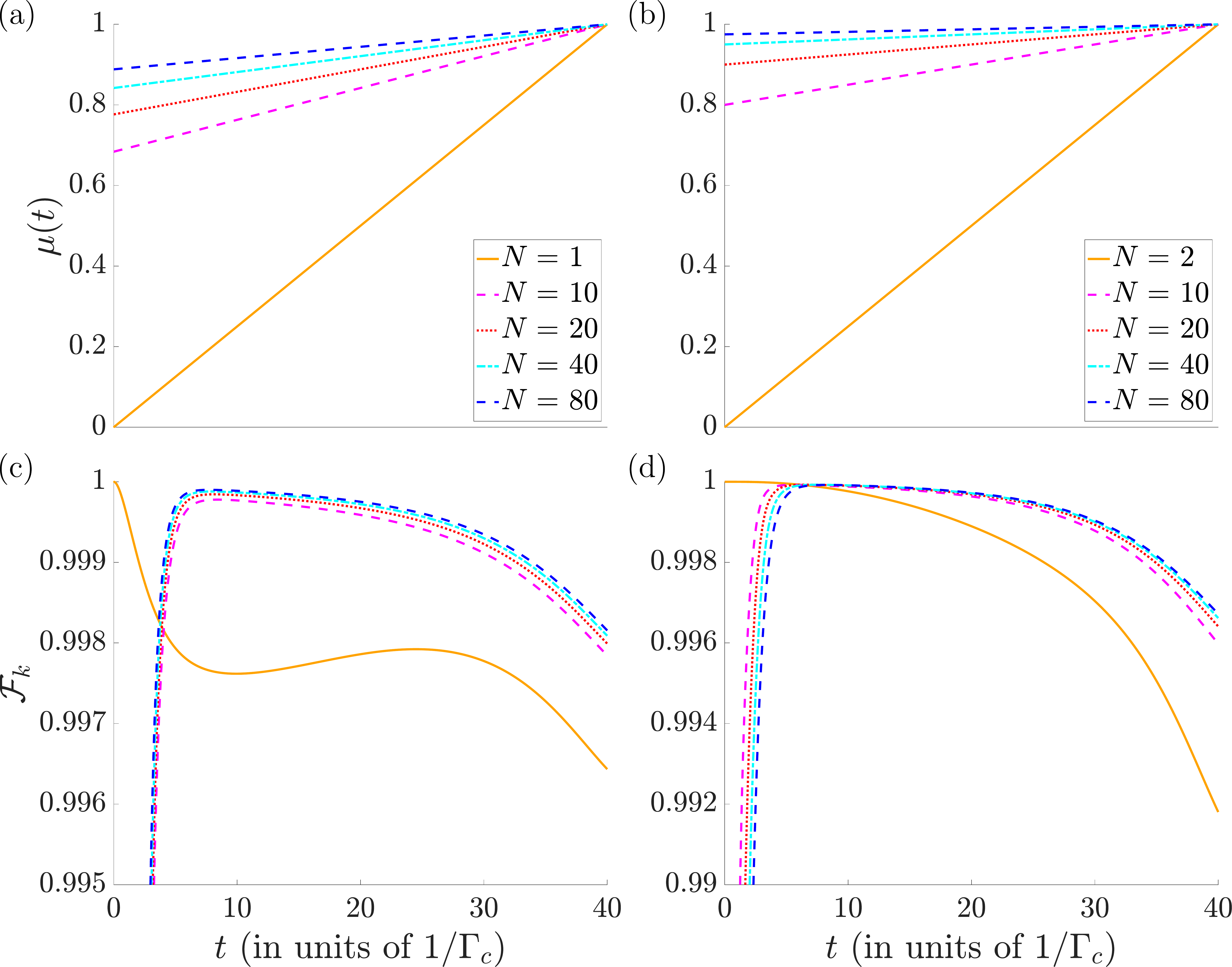}}
    \caption{(a-b) The ratio $\mu$ and (c-d) fidelity $\mathcal{F}_k$ using the quench scheme Eq.~\eqref{MuProfileQuench} with the desired state $k = 0$ for (a) and (c), while (b) and (d) show $k = N/2$.
    The parameters and curve colors are the same as in Fig.~\ref{N_loop_linear}.}
    \label{N_loop}
\end{figure}
We illustrate the advantage of our quenching scheme in Fig.~\ref{N_loop} where we choose $q = \sqrt{N}$ for the case $k = 0$ [Fig.~\ref{N_loop}(a) and (c)] and select $q = 2$ for $k = N/2$ [Fig.~\ref{N_loop}(b) and (d)] such that $\dot{\mu} \xi_{k,f} \sim 1/40$.
Therefore, as the atom number increases, we quench to a larger value of $\mu (t = 0) = 1 - 1/\sqrt{N}$ and then dynamically evolve $\mu$ with a lower slope so that the system's dynamics remain adiabatic, as shown in Figure~\ref{N_loop}(a) and (b).
At the quench $t = 0$, the fidelity may be calculated using $\abs{\ip{\psi_k (\mu = 0)}{\psi_k (\mu = 1 - q/N)}}^2 = \mathcal{N}_k^2 N!$ and Eq.~\eqref{N_k} which reveals a scaling that decreases as $N$ increases.
Meanwhile, Figs.~\ref{N_loop}(c) and (d) displays an enhancement of $F_{k,f}$ compared to Fig.~\ref{N_loop_linear} as all fidelities end above $0.99$ using the quench scheme.
Moreover, there is an enhancement of $F_{k,f}$ for large atom numbers compared to the single and two atom cases.
This enhancement grows slightly with $N$ due to the relative difference between $\max \left( \xi_k \right)$ and $\xi_{k,f}$ decreasing with increasing $N$, suggesting our choice of $\dot{\mu}$ becomes better when more atoms are in the system.
It must also be noted that the state reaches exactly $\mu = 1$ which is in contrast to the scheme proposed in~\cite{DallaTorre} where one quenches to $\mu = 1 - \varepsilon$, for a small constant $\varepsilon$, and then lets the system damp back into the desired state.
With this purely quench scheme, one can only achieve the $k = N/2$ state, must wait a exceedingly long time for the system to reach steady-state, and may not damp to exactly $\mu = 1$ as the system would equilibrate in an unpure, fully mixed state since the eigenstates are degenerate.
The study of how the quantum metrological usefulness of the state, i.e. the quantum Fisher information, varies with $\varepsilon$ is one of the subject of future work.

\begin{figure}
    \centerline{\includegraphics[width=\linewidth]{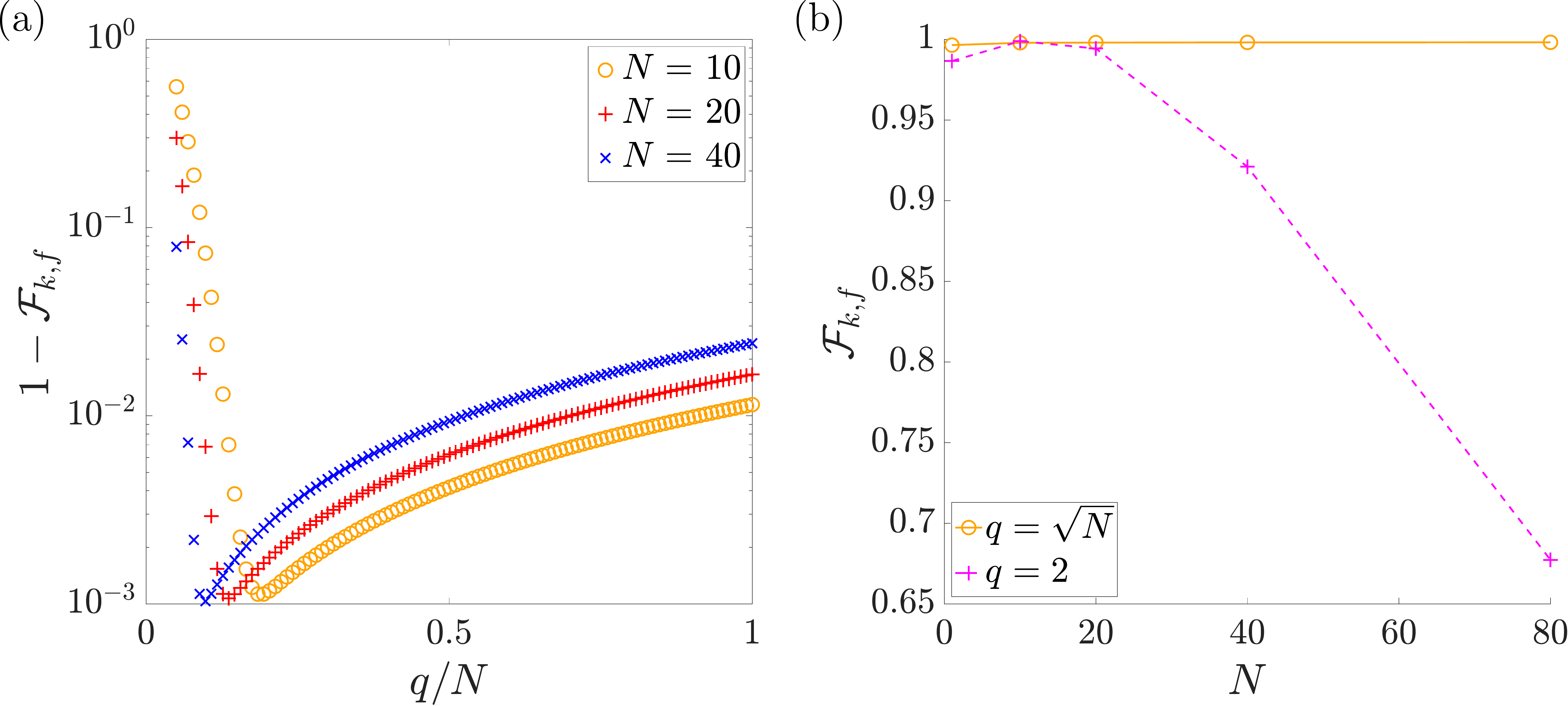}}
    \caption{(a) The difference of the final fidelity and unity $1 - \mathcal{F}_{k,f}$ as a function of the number $q/N$ for $N = 10$ (orange circles), $N = 20$ (red pluses), and $N = 40$ (blue crosses).
    This changes the value of $\mu (t = 0)$ and the slope of the linear ramp as we quench to different values.
    (b) The final fidelity as a function of atom number for the cases $q = \sqrt{N}$ (solid orange curve with circles) and $q = 2$ (dashed magenta curve with pluses).
    Both plots have $k = 0$, $t_f = 40/\Gamma_c$, and $\nu = 0$.}
    \label{N_loop_final_fidelity}
\end{figure}
Finally, we demonstrate in Fig.~\ref{N_loop_final_fidelity}(a) that one must make the correct choice of $q$ to achieve the desired dynamics.
Here, we consider the case $k = 0$ and examine the value of  $1 - \mathcal{F}_{k,f}$ as a function of $q$ for three different atom numbers.
For the points that we calculate, the maximum value of the final fidelity for the cases $N = 10$ (orange circles), $N = 20$ (red pluses), and $N = 40$ (blue crosses) is obtained at $q/N = 0.184$, $0.136$ and $0.098$, respectively.
We see that for $N = 10$, the maximum final fidelity is obtained when $q \approx 2$.
As $N$ increases, however, the maximum final fidelity occurs at a value of $q$ that approaches $\sqrt{N}$, which was the value used in Figs.~\ref{N_loop}(a) and (c).
We examine this behavior further in Fig.~\ref{N_loop_final_fidelity}(b) where we plot the final fidelity obtained with the choices $q = \sqrt{N}$ (solid orange curve with circles) and $q = 2$ (dashed pink curve with pluses) as a function of atom number.
We see that the original choice of $q = \sqrt{N}$ allows for a robust final fidelity that is extremely high $\mathcal{F}_{k,f} > 0.99$, while $\mathcal{F}_{k,f}$ in the $q = 2$ case drops off rather quickly with increasing $N$.
As shown in Figs.~\ref{N_loop}(a) and (b), $\mu$ in the $q = 2$ case quenches to a higher value before evolving with a more gradual slope as compared to $q = \sqrt{N}$. 
Therefore, we find for large $N$ that the $q = 2$ case cannot reach as high of a final fidelity as $q = \sqrt{N}$ even though its dynamics are ``more adiabatic.''
The reason for this is that process with $q = 2$ for the given $t_f$ does not have enough time to fully damp back into the DFS before $\xi_k$ spikes towards its final value and the dynamics become less adiabatic.
This further demonstrates why the choices of $q$ used in Fig.~\ref{N_loop} are (nearly) ideal, although this may be optimized further, for example by finding the actual value of $\max \left( \xi_k \right)$ to use for selecting the value of $\dot{\mu}$.
Furthermore, the choice of $t_f$ can also be varied to achieve either faster dynamics or higher final fidelities and a potential trade-off relation between these two objectives would be interesting to investigate, but we do not pursue this course of action here.

\section{Conclusion and Outlook} \label{ConclusionOutlook}
In this work, we proposed a protocol to adiabatically control a many-body system interacting with a highly dissipative cavity utilizing a DFS in the presence of collective decoherence.
We presented a method to analytically obtain the eigenstates of a non-Hermitian quadratic jump operator utilizing the Schwinger boson representation.
We then used the criterion for a dynamically stable DFS Eq.~\eqref{DFScondition} to derive a cavity driving profile which deconstructively interferes with the atomic ensemble's emission amplitude for a given DFS eigenstate.
This allowed us to engineer a desired $\hat{J}^x$ eigenstate by adiabatically following the time evolution of a DFS eigenstate as the classical driving fields are varied, which we demonstrated using a linear increase of the ratio $\mu$ from $0$ to $1$.
We then investigated how quickly one may evolve the parameters of the system by studying the adiabaticity parameter of the system.
Here, we found that for $N \gg 1$, one may vary $\mu$ extremely rapidly for a majority of the process before the adiabaticity parameter drastically increases towards a large final value such that the evolution of $\mu$ needs to be gradual for the remainder of the process.
This motivated the introduction of a quench scheme in which we quench to a value of $\mu = 1 - q/N$ and then evolve the system gradually for the remainder of the process with a slope that is modified for different atom numbers.
This scheme had the ability to adiabatically construct the desired states with extremely high final fidelity $\mathcal{F}_{k,f} > 0.99$ and we showed that $\mathcal{F}_{k,f}$ increases with $N$.
We concluded by investigating the optimal value of $q$ to maximize the final fidelity for $N \gg 10$.
A more complicated driving profile of $\mu$ may allow for even more optimized dynamics, but we did not pursue this prospect in this work.

In our analysis, we have neglected single atom spontaneous emission from the excited states $\ket{l}$ and $\ket{r}$ based on large detuning $\Delta_l, \Delta_r \gg \Omega_1, \Omega_2, \sqrt{N} g$.
While the use of stimulated Raman transitions allows one to engineer an artificial linewidth of the spins, decreasing the single atom linewidth will also decrease the collective emission rate $\Gamma_c$.
Therefore, an important experimental consideration is the single atom cooperativity parameter $\mathcal{C}$ which must be large in order to ignore single atom emission throughout the timescales used in Fig.~\ref{N_loop}.

To overcome the requirement of a large $\mathcal{C}$ cavity, one can instead create a scheme that drives the system with dynamics that needs not be adiabatic.
Therefore, a natural next step is to develop an adiabatic shortcut to engineer an adiabtic shortcut which can create a desired DFS eigenstate with  arbitrarily fast evolution time so long as one is provided with arbitrary large driving intensities.
Another requirement is that the adiabatic drive must have enough time to begin to follow the $k^{\text{th}}$ eigenstate as the eigenestates are degenerate at the beginning of the process.
With the addition of a second classical drive of the cavity $\eta_s$, one can superadiabatically create the spin coherent states $k = 0, N$ using the shortcut protocol developed in~\cite{Wu}, but this fails to purely create the ring states.
This is because the $k^{\text{th}}$ DFS eigenstate evolves with a certain amount of overlap with the two neighboring eigenstates $k \pm 1$ [see Eq.~\eqref{TimeDerivative}], and the shortcut drive is only able to cancel out the overlap with one of these neighboring states (determined by the sign of $\eta_s$) while it, in fact, enhances the overlap with the other neighboring state.
The spin coherent states are able to be created with $\eta_s$ as each only has one neighboring eigenstate.
To overcome this limitation, one might introduce an extra degree of freedom to the system in order to remove the coupling to both neighboring eigenstates and thus create every eigenstate, including the spin squeezed $k = N/2$ state, with dynamics that does not need to be adiabatic.

\section*{Acknowledgements}
We would like to thank John D. Wilson, Ana Maria Rey, and Graeme Smith for useful discussions.
This research was supported
by the NSF PFC Grant No. 1734006; NSF AMO Grant No. 1806827; and the NSF Q-SEnSE Grant No. OMA 2016244.

\bibliography{references.bib}

\appendix

\section{Derivation of the Effective Master Equation} \label{MasterEquationDerivation}

\subsection{Two-level Hamiltonian}
We consider the cavity and four-level atom interaction shown in Fig.~\ref{Model}, which has the many-body Hamiltonian in the Schr\"odinger picture given by
\begin{equation} \label{SchrodingerHamiltonian}
    \begin{aligned}
\hat{H}_0 =& \sum_{j = 1}^N \frac{\hbar \omega_{\uparrow}}{2} \left( \ket{\uparrow}_j \bra{\uparrow}_j - \ket{\downarrow}_j \bra{\downarrow}_j \right) + \hbar \omega_l \ket{l}_j \bra{l}_j \\
&+ \hbar \omega_r \ket{r}_j \bra{r}_j + \hbar \omega_c \hat{a}^{\dagger} \hat{a} + \hbar \left( \eta \hat{a}^{\dagger} e^{-i \omega_d t} + \text{H.c.} \right)\\
&+ \hbar g \left[ \left( \ket{l}_j \bra{\downarrow}_j \hat{a} + \text{H.c.} \right) + \left( \ket{r}_j \bra{\uparrow}_j \hat{a} + \text{H.c.} \right) \right] \\
&+ \frac{\hbar \Omega_1}{2} \left( \ket{l}_j \bra{\uparrow}_j e^{-i \omega_1 t} + \text{H.c.} \right) \\
&+ \frac{\hbar \Omega_2}{2} \left( \ket{r}_j \bra{\downarrow}_j e^{-i \omega_2 t} + \text{H.c.} \right),
    \end{aligned}
\end{equation}
where we have set the zero energy half way between $\ket{\uparrow}$ and $\ket{\downarrow}$.
Here, we have defined the annihilation (creation) operator $\hat{a}$ ($\hat{a}^{\dagger}$) of a cavity mode with frequency $\omega_c$, and the bare frequencies $\omega_{\uparrow}, \omega_{l}, \omega_{r}$ of the three states $\ket{\uparrow}, \ket{l}, \ket{r}$, respectively, with respect to the frequency of $\ket{\downarrow}$.
We have also assumed that spontaneous decay from $\ket{l}$ and $\ket{r}$ is negligible, $\gamma_l \approx \gamma_r \approx 0$, and that the laser field that drives the cavity with amplitude $\eta$ has frequency $\omega_d$.
We then move into the interaction picture that induces the rotation $\hat{\rho} \to \tilde{\hat{\rho}} = \hat{U} \hat{\rho} \hat{U}^{\dagger}$  with $\hat{U} = \exp \left[ i \hat{H}' t/ \hbar \right]$.
Defining $\hat{J}^z = \sum_{j = 1}^N \left( \ket{\uparrow}_j \bra{\uparrow}_j - \ket{\downarrow}_j \bra{\downarrow}_j \right)/2$, we set
\begin{equation}
    \begin{aligned} 
\hat{H}' = & \frac{\hbar \left( \omega_1 - \omega_2 \right)}{2} \hat{J}^z + \sum_{j = 1}^N \hbar \omega_2 \ket{r}_j \bra{r}_j \\
&+ \frac{\hbar \left( \omega_1 + \omega_2 \right)}{2} \left( \ket{l}_j \bra{l}_j + \hat{a}^{\dagger} \hat{a} \right),
    \end{aligned}
\end{equation}
such that the Hamiltonian becomes
\begin{equation}
    \begin{aligned}
\hat{H}_I =& \hbar \Delta_{\uparrow} \hat{J}^z - \hbar \Delta_c \hat{a}^{\dagger} \hat{a} + \hbar \left( \eta \hat{a}^{\dagger} e^{-i \Delta_d t} + \text{H.c.} \right) \\ 
&- \sum_{j = 1}^N \hbar \Delta_l \ket{l}_j \bra{l}_j - \hbar \Delta_r \ket{r}_j \bra{r}_j \\ 
&+ \hbar g \left[ \left( \ket{l}_j \bra{\downarrow}_j \hat{a} + \text{H.c.} \right) + \left( \ket{r}_j \bra{\uparrow}_j \hat{a} + \text{H.c.} \right) \right] \\
&+ \frac{\hbar \Omega_1}{2} \left( \ket{l}_j \bra{\uparrow}_j + \text{H.c.} \right) + \frac{\hbar \Omega_2}{2} \left( \ket{r}_j \bra{\downarrow}_j + \text{H.c.} \right),
    \end{aligned}
\end{equation}
where we have introduced the detunings $\Delta_{\uparrow} = \omega_{\uparrow} - \left( \omega_1 - \omega_2 \right)/2$, $\Delta_c = \left( \omega_1 + \omega_2 \right)/2 - \omega_c$, $\Delta_l = \left( \omega_1 + \omega_2 \right)/2 - \omega_l$, $\Delta_r = \omega_2 - \omega_r$, and $\Delta_d = \omega_d - \left( \omega_1 + \omega_2 \right)/2$.

We now assume that the detunings of the laser fields are very large $\abs{\Delta_l}, \abs{\Delta_r} \gg \Omega_1, \Omega_2, \sqrt{N} g$ to adiabatically eliminate the exited states $\ket{l}$ and $\ket{r}$ over a coarse-grained timescale~\cite{Steck}.
We obtain
\begin{equation} \label{NBodyHamiltonianStark}
    \begin{aligned}
\hat{H} =& -\hbar \Delta_c \hat{a}^{\dagger} \hat{a} + \hbar \Delta_{\uparrow} \hat{J}^z + \hbar \left( \eta \hat{a}^{\dagger} e^{-i \Delta_d t} + h.c. \right) \\
&+ \frac{\hbar g}{2} \left[ \frac{\Omega_1}{\Delta_l} \left( \hat{J}^- \hat{a}^{\dagger} + h.c. \right) + \frac{\Omega_2}{\Delta_r} \left( \hat{J}^+ \hat{a}^{\dagger} + h.c. \right) \right] \\
&+ \sum_{j=1}^N \left( \frac{\hbar \Omega_1^2}{4 \Delta_l} \ket{\uparrow}_j \bra{\uparrow}_j + \frac{\hbar \Omega_2^2}{4 \Delta_r} \ket{\downarrow}_j \bra{\downarrow}_j \right. \\
&\left. + \left[ \frac{\hbar g^2}{\Delta_r} \ket{\uparrow}_j \bra{\uparrow}_j + \frac{\hbar g^2}{\Delta_l} \ket{\downarrow}_j \bra{\downarrow}_j \right] \hat{a}^{\dagger} \hat{a} \right).
    \end{aligned} 
\end{equation}
We may now set $\Delta_{\uparrow} = \Delta_d = 0$.
Moreover, the last two lines of Eq.~\eqref{NBodyHamiltonianStark} represents the AC Stark shifts which can be ignored with physical justifications.
For example, if the ground states are hyperfine split states, an time-dependent external magnetic field can shift the levels in such a way to compensate for the Stark shifts proportional to $\Omega_1^2$ and $\Omega_2^2$, while the Stark shifts proportional to $g^2$ can be neglected because the cavity mode decays on an extremely fast timescale, as discussed Appendix~\ref{CavityModeElimination}, and thus $\expval{\hat{a}^{\dagger} \hat{a}} \approx 0$.
We therefore obtain the final effective two-level Hamiltonian:
\begin{equation} \label{NBodyHamiltonian}
    \begin{aligned}
\hat{H} =& -\hbar \Delta_c \hat{a}^{\dagger} \hat{a} + \hbar \left( \eta \hat{a}^{\dagger} + \text{\text{H.c.}} \right) \\
&+ \frac{\hbar g}{2} \left[ \frac{\Omega_1}{\Delta_l} \hat{J}^- \hat{a}^{\dagger} + \frac{\Omega_2}{\Delta_r} \hat{J}^+ \hat{a}^{\dagger} + \text{\text{H.c.}} \right].
    \end{aligned}
\end{equation}
In addition, we introduce dissipation of the cavity mode using the Lindblad superoperator
Eq.~\eqref{LindbladSuperoperator} with jump operator $\hat{L}_{\text{cav}} = \sqrt{2 \kappa} \hat{a}$.

\subsection{Rotating Frame} \label{RotatingFrame}
To simplify the calculation of the final Hamiltonian, we assume the classical fields take the form
\begin{equation}
    \Omega_1 = \abs{\Omega_1} e^{-i \phi_1}, \quad \Omega_2 = \abs{\Omega_2} e^{-i \phi_2},
\end{equation}
so that Eq.~\eqref{NBodyHamiltonian} becomes
\begin{equation}
    \begin{aligned}
\hat{H} = & -\hbar \Delta_c \hat{a}^{\dagger} \hat{a} + \hbar \left( \eta \hat{a}^{\dagger} + \text{\text{H.c.}} \right) \\
& + \frac{\hbar g}{2} \left[ \hat{a}^{\dagger} \left( \frac{\abs{\Omega_1}}{\Delta_l} e^{-i \phi_1} \hat{J}^- + \frac{\abs{\Omega_2}}{\Delta_r} e^{-i \phi_2} \hat{J}^+ \right) + \text{\text{H.c.}} \right].
    \end{aligned}
\end{equation}
We now make rotations of the quantization axes of the collective dipole and the cavity in order to cancel the phases in $\hat{H}$.
We therefore make the choices 
\begin{equation}
    \hat{a}^{\dagger} \rightarrow \hat{a}^{\dagger} e^{i \phi_a}, \quad \hat{J}^+ \rightarrow \hat{J}^+ e^{-i \phi_J},
\end{equation}
with the phases
\begin{equation}
    \phi_a = \frac{1}{2} \left( \phi_1 + \phi_2 \right), \quad \phi_J = \frac{1}{2} \left( \phi_1 - \phi_2 \right),
\end{equation}
so that we have
\begin{equation}
    \tilde{\hat{L}}_{\text{cav}} = \sqrt{2 \kappa} \hat{a} e^{- i \phi_a},
\end{equation}
as well as
\begin{equation}
    \begin{aligned} 
\tilde{\hat{H}} = & -\hbar \Delta_c \hat{a}^{\dagger} \hat{a} \\
& + \hbar \left[ \hat{a}^{\dagger} \left( \eta + \frac{g \abs{\Omega_1}}{2 \Delta_l} \hat{J}^- + \frac{g \abs{\Omega_2}}{2 \Delta_r} \hat{J}^+ \right) + \text{\text{H.c.}} \right].
    \end{aligned}
\end{equation}
Here, we have rotated the pump frequency
\begin{equation} \label{RotatedEta}
    \eta \rightarrow \eta e^{i \phi_a},
\end{equation}
such that the phases cancel in the final form of our Hamiltonian.

\subsection{Elimination of Dissipative Cavity Mode} \label{CavityModeElimination}
We now assume that the cavity mode $\hat{a}$ decays rapidly so that is in the bad cavity limit, meaning $\kappa \gg \sqrt{N} \eta, \sqrt{N} g \abs{\Omega_1}/(2 \abs{\Delta_l}), \sqrt{N} g \abs{\Omega_2}/(2 \abs{\Delta_r})$.
In this limit, the cavity mode can be adiabatically eliminated so we may find an effective master equation for the atomic degrees of freedom.
We do this by projecting the system onto the vacuum state of the cavity mode and including effects from the atomic evolution and atom-cavity interaction only up to second order.
The resulting master equation for the reduced density operator $\hat{\rho}_{\text{at}} = \Tr_F \left[ \hat{\rho} \right]$, where $\Tr_F \left[ \cdot \right]$ is the partial trace over the cavity degrees of freedom, is given by 
\begin{equation} 
    \pd{\hat{\rho}_{\text{at}}} = \hat{\mathcal{L}}_{\text{at}} \hat{\rho}_{\text{at}} := \frac{1}{i\hbar} \left[ \hat{H}_{\text{at}}, \hat{\rho}_{\text{at}} \right] + \hat{\mathcal{D}} \left[ \hat{L} \right] \hat{\rho}_{\text{at}},
\end{equation}
with jump operator
\begin{equation} 
    \hat{L} = \sqrt{\frac{\kappa g^2 \abs{\Omega_1}^2}{2 \abs{\Delta_l}^2 \left( \Delta_c^2 + \kappa^2 \right)}} \left( \hat{J}^- + \abs{\frac{\Omega_2 \Delta_l}{\Omega_1 \Delta_r}} \hat{J}^+ + \frac{2 \eta \abs{\Delta_l}}{g \abs{\Omega_1}} \hat{\mathbb{I}} \right),
\end{equation}
and an effective Hamiltonian
\begin{equation} 
    \hat{H}_{\text{at}} = \frac{\hbar \Delta_c}{2 \kappa} \hat{L}^{\dagger} \hat{L}.
\end{equation}

\section{Calculation of Overlaps} \label{OverlapCalculation}

\subsection{Overlap of Eigenstates}
We now wish to derive a general formula for the overlap of two DFS eigenstates $\ket{\psi_k}$ and $\ket{\psi_{k'}}$. 
To do this, we assume that $\hat{c}_2^{\dagger}$ can be decomposed into a term proportional to $\hat{c}_1^{\dagger}$ and a complementary term $\hat{c}_{\perp}^{\dagger}$ such that
\begin{equation}
    \hat{c}_2^{\dagger} = a_1 \hat{c}_1^{\dagger} + a_{\perp} \hat{c}_{\perp}^{\dagger},
\end{equation}
where we have $\left[ \hat{c}_1, \hat{c}_{\perp}^{\dagger} \right] = 0$ by construction.
Therefore, the overlap between two eigenstates becomes
\begin{equation}
    \begin{aligned}
\ip{\psi_{k'}}{\psi_k} = & \mathcal{N}_{k'}^* \mathcal{N}_k \bra{0} \left( \hat{c}_1 \right)^{N - k'} \left( \hat{c}_2 \right)^{k'} \left( \hat{c}_1^{\dagger} \right)^{N - k} \left( \hat{c}_2^{\dagger} \right)^k \ket{0} \\
= & \mathcal{N}_{k'}^* \mathcal{N}_k \bra{0} \left( \hat{c}_1 \right)^{N - k'} \left( a_1^* \hat{c}_1 + a_{\perp}^* \hat{c}_{\perp} \right)^{k'} \times \\
& \left( \hat{c}_1^{\dagger} \right)^{N - k} \left( a_1 \hat{c}_1^{\dagger} + a_{\perp} \hat{c}_{\perp}^{\dagger} \right)^k \ket{0},
    \end{aligned}
\end{equation}
and since $\left[ \hat{c}_1, \hat{c}_{\perp} \right] = \left[ \hat{c}_1^{\dagger}, \hat{c}_{\perp}^{\dagger} \right] = 0$, we can use binomial theorem to expand
\begin{equation}
    \begin{aligned}
\ip{\psi_{k'}}{\psi_k} = & \mathcal{N}_{k'}^* \mathcal{N}_k \sum_{i=0}^{k'} \sum_{j=0}^k \binom{k'}{i} \binom{k}{j} {a_1^*}^i a_1^j {a_{\perp}^*}^{k'-i} a_{\perp}^{k-j} \times \\
& \bra{0} \left( \hat{c}_1 \right)^{N-k'+i} \left( \hat{c}_{\perp} \right)^{k'-i} \left( \hat{c}_1^{\dagger} \right)^{N-k+j} \left( \hat{c}_{\perp}^{\dagger} \right)^{k-j} \ket{0}.
    \end{aligned}
\end{equation}
We note that when $k'-i > k-j$ or $k'-i < k-j$, we obtain $\ip{\psi_{k'}}{\psi_k} = 0$.
Thus, we need $k'-i = k-j$ and so we set $j = k-k'+i$ to write 
\begin{equation}
    \begin{aligned} 
\ip{\psi_{k'}}{\psi_k} = & \mathcal{N}_{k'}^* \mathcal{N}_k \sum_{i=0}^{k'} \binom{k'}{i} \binom{k}{k-k'+i} \abs{a_1}^{2i} {a_1^*}^{k-k'} \times \\
& \abs{a_{\perp}}^{2 \left( k'-i \right)} \left( N - k' + i \right)! \left[ \hat{c}_1, \hat{c}_1^{\dagger} \right]^{N-k'+i} \times \\
& \left( k' - i \right)! \left[ \hat{c}_{\perp}, \hat{c}_{\perp}^{\dagger} \right]^{k'-i}.
    \end{aligned}
\end{equation}

We now must calculate
\begin{equation}
    \left[ \hat{c}_1, \hat{c}_2^{\dagger} \right] = \left[ \hat{c}_1, a_1 \hat{c}_1^{\dagger} + a_{\perp} \hat{c}_{\perp}^{\dagger} \right] = a_1 \left[ \hat{c}_1, \hat{c}_1^{\dagger} \right],
\end{equation}
and use $\left[ \hat{c}_1, \hat{c}_1^{\dagger} \right] = 1 + \mu^2$ and $\left[ \hat{c}_1, \hat{c}_2^{\dagger} \right] = 1 - \mu^2$ such that
\begin{equation}
    a_1 = \frac{1 - \mu^2}{1 + \mu^2}.
\end{equation}
To find the other coefficient, we first define the complementary operator as
\begin{equation}
    \hat{c}_{\perp}^{\dagger} = \hat{b}_{\uparrow}^{\dagger} - \mu \hat{b}_{\downarrow}^{\dagger},
\end{equation}
so that we have $\left[ \hat{c}_{\perp}, \hat{c}_{\perp}^{\dagger} \right] = 1 + \mu^2$ and $\left[ \hat{c}_{\perp}, \hat{c}_2^{\dagger} \right] = -2 \mu$.
Calculating
\begin{equation}
    \left[ \hat{c}_{\perp}, \hat{c}_2^{\dagger} \right] = \left[ \hat{c}_{\perp}, a_1 \hat{c}_1^{\dagger} + a_{\perp} \hat{c}_{\perp}^{\dagger} \right] = a_{\perp} \left[ \hat{c}_{\perp}, \hat{c}_{\perp}^{\dagger} \right],
\end{equation}
we find
\begin{equation}
    a_{\perp} = - \frac{2 \mu}{1 + \mu^2}.
\end{equation}
Therefore, we can write the general overlap as
\begin{equation} \label{EigenstateOverlaps}
    \begin{aligned} 
\ip{\psi_{k'}}{\psi_k} = & \mathcal{N}_{k'}^* \mathcal{N}_k \left( 1 + \mu^2 \right)^N \sum_{i=0}^{k'} \binom{k'}{i} \binom{k}{k-k'+i} \times \\
& \abs{a_1}^{2i} {a_1^*}^{k-k'} \abs{a_{\perp}}^{2 \left( k'-i \right)} \left( N - k' + i \right)! \left( k' - i \right)!.
    \end{aligned}
\end{equation}
Using this with $k=k'$, we can derive an analytic form of the normalization factor:
\begin{equation} \label{N_k}
    \frac{1}{\mathcal{N}_k} = \sqrt{\sum_{i=0}^k \frac{\binom{k}{i} k! \left( N-k+i \right)! \left( 1 - \mu^2 \right)^{2i} \left( 4 \mu^2 \right)^{k-i}}{i! \left( 1 + \mu^2 \right)^{2k -N}}},
\end{equation}
when $\mu \neq 1$.

\subsection{Overlap of Complementary States}
We can perform a similar calculation for the overlap between two complementary states $\ket{\psi_n^{\perp}}$ and $\ket{\psi_{n'}^{\perp}}$ by assuming $\hat{d}_2^{\dagger}$ can be decomposed into a term proportional to $\hat{d}_1^{\dagger}$ and a complementary term $\hat{d}_{\perp}^{\dagger}$ such that
\begin{equation}
    \hat{d}_2^{\dagger} = b_1 \hat{d}_1^{\dagger} + b_{\perp} \hat{d}_{\perp}^{\dagger},
\end{equation}
where we have $\left[ \hat{d}_1, \hat{d}_{\perp}^{\dagger} \right] = 0$ by construction.
Therefore, the overlap between two complementary states becomes
\begin{equation}
    \begin{aligned}
\ip{\psi_{n'}^{\perp}}{\psi_n^{\perp}} = & \left( 2 \mu \right)^{2 N} {\mathcal{N}_{n'}^{\perp}}^* \mathcal{N}_n^{\perp} \sum_{i=0}^{n'} \sum_{j=0}^n \binom{n'}{i} \binom{n}{j} {b_1^*}^i b_1^j \times \\
& {b_{\perp}^*}^{n'-i} b_{\perp}^{n-j} \bra{0} \left( \hat{d}_1 \right)^{N-n'+i} \left( \hat{d}_{\perp} \right)^{n'-i} \times \\
& \left( \hat{d}_1^{\dagger} \right)^{N-n+j} \left( \hat{d}_{\perp}^{\dagger} \right)^{n-j} \ket{0},
    \end{aligned}
\end{equation}
where we have used binomial theorem since $\left[ \hat{d}_1, \hat{d}_{\perp} \right] = \left[ \hat{d}_1^{\dagger}, \hat{d}_{\perp}^{\dagger} \right] = 0$.
We again find that $\ip{\psi_{n'}^{\perp}}{\psi_n^{\perp}} = 0$ when $n'-i > n-j$ or $n'-i < n-j$, such that we need $j = n-n'+i$.
Therefore, we find
\begin{equation}
    \begin{aligned} 
\ip{\psi_{n'}^{\perp}}{\psi_n^{\perp}} = & \left( 2 \mu \right)^{2 N} {\mathcal{N}_{n'}^{\perp}}^* \mathcal{N}_n^{\perp} \sum_{i=0}^{n'} \binom{n'}{i} \binom{n}{n-n'+i} \times \\ 
& \abs{b_1}^{2i} {b_1^*}^{n-n'} \abs{b_{\perp}}^{2 \left( n'-i \right)} \left( N - n' + i \right)! \times \\
& \left[ \hat{d}_1, \hat{d}_1^{\dagger} \right]^{N-n'+i} \left( n' - i \right)! \left[ \hat{d}_{\perp}, \hat{d}_{\perp}^{\dagger} \right]^{n'-i}.
    \end{aligned}
\end{equation}

We now must calculate
\begin{equation}
    \left[ \hat{d}_1, \hat{d}_2^{\dagger} \right] = \left[ \hat{d}_1, b_1 \hat{d}_1^{\dagger} + b_{\perp} \hat{d}_{\perp}^{\dagger} \right] = b_1 \left[ \hat{d}_1, \hat{d}_1^{\dagger} \right],
\end{equation}
and use $\left[ \hat{d}_1, \hat{d}_1^{\dagger} \right] = \left( \mu^2 + 1 \right)/(4 \mu^2)$ and $\left[ \hat{d}_1, \hat{d}_2^{\dagger} \right] = \left( \mu^2 - 1 \right)/(4 \mu^2)$ such that
\begin{equation}
    b_1 = \frac{\mu^2 - 1}{\mu^2 + 1}.
\end{equation}
To find the other coefficient, we first define the complementary operator as
\begin{equation}
    \hat{d}_{\perp} = \frac{1}{2} \hat{b}_{\uparrow} - \frac{1}{2 \mu} \hat{b}_{\downarrow},
\end{equation}
so that we have $\left[ \hat{d}_{\perp}, \hat{d}_{\perp}^{\dagger} \right] = \left( \mu^2 + 1 \right)/(4 \mu^2)$ and $\left[ \hat{d}_{\perp}, \hat{d}_2^{\dagger} \right] = - 1/(2 \mu)$.
Calculating
\begin{equation}
    \left[ \hat{d}_{\perp}, \hat{d}_2^{\dagger} \right] = \left[ \hat{d}_{\perp}, b_1 \hat{d}_1^{\dagger} + b_{\perp} \hat{d}_{\perp}^{\dagger} \right] = b_{\perp} \left[ \hat{d}_{\perp}, \hat{d}_{\perp}^{\dagger} \right],
\end{equation}
we find
\begin{equation}
    b_{\perp} = - \frac{2 \mu}{\mu^2 + 1}.
\end{equation}
Therefore, we can write the overlap of complementary states as
\begin{equation} \label{ComplementaryStateOverlaps}
    \begin{aligned} 
\ip{\psi_{n'}^{\perp}}{\psi_n^{\perp}} = & \left( \mu^2 + 1 \right)^N {\mathcal{N}_{n'}^{\perp}}^* \mathcal{N}_n^{\perp} \sum_{i=0}^{n'} \binom{n'}{i} \binom{n}{n-n'+i} \times \\ 
& \abs{b_1}^{2i} {b_1^*}^{n-n'} \abs{b_{\perp}}^{2 \left( n'-i \right)} \left( N - n' + i \right)! \left( n' - i \right)!.
    \end{aligned}
\end{equation}
Setting $n=n'$, we can find the normalization factor to be 
\begin{equation} \label{N_n_perp}
    \frac{1}{\mathcal{N}_n^{\perp}} = \sqrt{\sum_{i=0}^n \frac{\binom{n}{i} n! \left( N-n+i \right)! \left( \mu^2 - 1 \right)^{2i} \left( 4 \mu^2 \right)^{n-i}}{i! \left( \mu^2 +1 \right)^{2n - N}}}.
\end{equation}

\section{Calculation of the Adiabaticity Parameter} \label{AdiabaticParameterCalculation}

\subsection{General Form}
In order to quantify how quickly the system's drives can vary while remaining in the adiabatic regime, we now calculate the adiabaticity parameter 
\begin{equation}
    \Xi_k = \max_{n} \abs{\frac{4 \ip{\psi_n^{\perp}}{\rpd \psi_k}}{\alpha_{nk} + i \zeta_n}} \ll 1, \quad \forall n \neq k.
\end{equation}
We first find the time derivative using Eq.~\eqref{DFSEigenstates}:
\begin{equation} \label{TimeDerivative}
    \begin{aligned} 
\rpd \ket{\psi_k} = & \frac{\rpd \mathcal{N}_k}{\mathcal{N}_k} \ket{\psi_k} + \mathcal{N}_k \left[ (N-k) \left( \rpd \hat{c}_1^{\dagger} \right) \left( \hat{c}_1^{\dagger} \right)^{N-k-1} \times \right. \\ 
& \left. \left( \hat{c}_2^{\dagger} \right)^k + k \left( \rpd \hat{c}_2^{\dagger} \right) \left( \hat{c}_1^{\dagger} \right)^{N-k} \left( \hat{c}_2^{\dagger} \right)^{k-1} \right] \ket{0} \\
= & \frac{\rpd \mathcal{N}_k}{\mathcal{N}_k} \ket{\psi_k} + \dot{\mu} \hat{b}_{\uparrow}^{\dagger} \mathcal{N}_k \left[ (N-k) \left( \hat{c}_1^{\dagger} \right)^{N-k-1} \left( \hat{c}_2^{\dagger} \right)^k \right. \\
& \left. - k \left( \hat{c}_1^{\dagger} \right)^{N-k} \left( \hat{c}_2^{\dagger} \right)^{k-1} \right] \ket{0} \\
= & \frac{\rpd \mathcal{N}_k}{\mathcal{N}_k} \ket{\psi_k} + \frac{\dot{\mu}}{2 \mu} \mathcal{N}_k \left[ \frac{N}{\mathcal{N}_k} \ket{\psi_k} \right. \\
& \left. - \frac{N-k}{\mathcal{N}_{k+1}} \ket{\psi_{k+1}} - \frac{k}{\mathcal{N}_{k-1}} \ket{\psi_{k-1}} \right],
    \end{aligned}
\end{equation}
where we have used
\begin{equation}
    \hat{b}_{\uparrow}^{\dagger} = \frac{1}{2 \mu} \left( \hat{c}_1^{\dagger} - \hat{c}_2^{\dagger} \right).
\end{equation}
We note that we can write 
\begin{equation}
    \hat{J}^z = \frac{1}{2} \left( \hat{b}_{\uparrow}^{\dagger} \hat{b}_{\uparrow} - \hat{b}_{\downarrow}^{\dagger} \hat{b}_{\downarrow} \right) = - \frac{1}{2} \left( \hat{d}_2^{\dagger} \hat{c}_1 + \hat{d}_1^{\dagger} \hat{c}_2 \right),
\end{equation}
such that
\begin{equation} \label{Jz_on_eigenstate}
    \begin{aligned}
\hat{J}^z \ket{\psi_k} &= - \frac{\mathcal{N}_k}{2} \left[ \hat{d}_2^{\dagger} \hat{c}_1 + \hat{d}_1^{\dagger} \hat{c}_2 \right] \left( \hat{c}_1^{\dagger} \right)^{N-k} \left( \hat{c}_2^{\dagger} \right)^k \ket{0} \\
&= - \frac{\mathcal{N}_k}{2} \left[ \frac{N-k}{\mathcal{N}_{k+1}^{\perp}} \ket{\psi_{k+1}^{\perp}} + \frac{k}{\mathcal{N}_{k-1}^{\perp}} \ket{\psi_{k-1}^{\perp}} \right],
    \end{aligned}
\end{equation}
so that we may relate
\begin{equation}
    \rpd \ket{\psi_k} = \left[ \frac{\rpd \mathcal{N}_k}{\mathcal{N}_k} + \frac{\dot{\mu}}{2 \mu} \left( N + 2 \hat{J}^z \right) \right] \ket{\psi_k}.
\end{equation}
Projecting a complementary state on the left, we find
\begin{equation}
    \begin{aligned} 
\ip{\psi_n^{\perp}}{\rpd \psi_k} = & \frac{\dot{\mu}}{\mu} \bra{\psi_n^{\perp}} \hat{J}^z \ket{\psi_k} \\
= & - \frac{\dot{\mu}}{2 \mu} \mathcal{N}_k \left[ \frac{N-k}{\mathcal{N}_{k+1}} \ip{\psi_{k+1}^{\perp}}{\psi_{k+1}} \delta_{n,k+1} \right. \\
& \left. + \frac{k}{\mathcal{N}_{k-1}} \ip{\psi_{k-1}^{\perp}}{\psi_{k-1}} \delta_{n,k-1} \right],
    \end{aligned}
\end{equation}
which we combine with Eq.~\eqref{PerpOverlap} to obtain
\begin{equation} \label{FullDerivativeOverlap}
    \begin{aligned} 
\ip{\psi_n^{\perp}}{\rpd \psi_k} = & - \dot{\mu} \left( 2 \mu \right)^{N-1} \mathcal{N}_k \left[ \mathcal{N}_{k+1}^{\perp} (N-k)! (k+1)! \delta_{n,k+1} \right. \\
& \left. + \mathcal{N}_{k-1}^{\perp} (N-k+1)! k! \delta_{n,k-1} \right].
    \end{aligned}
\end{equation}
We therefore only have to consider $n = k \pm 1$ when calculating the adiabaticity criteria
\begin{equation}
    \Xi_k = \max_{n = k \pm 1} \abs{\frac{4 \ip{\psi_n^{\perp}}{\rpd \psi_k}}{\alpha_{nk} + i \zeta_n}} \ll 1.
\end{equation}

To calculate the terms in the denominator, we first show that the complementary states $\ket{\psi_n^{\perp}}$ are right eigenstates of $\tilde{L}^{\dagger}$,
\begin{equation}
    \begin{aligned} 
\tilde{L}^{\dagger} \ket{\psi_n^{\perp}} = & \left( 2 \mu \right)^N \mu \mathcal{N}_{n}^{\perp} \left( \hat{d}_1^{\dagger} \hat{c}_1 - \hat{d}_2^{\dagger} \hat{c}_2 \right) \left( \hat{d}_1^{\dagger} \right)^{N-n} \left( \hat{d}_2^{\dagger} \right)^n \ket{0} \\
= & \left( 2 \mu \right)^N \mu \mathcal{N}_{n}^{\perp} \left( \hat{d}_1^{\dagger} \left[ \left( \hat{d}_1^{\dagger} \right)^{N-n} \hat{c}_1 \right. \right. \\
& \left. \left. + (N-n) \left( \hat{d}_1^{\dagger} \right)^{N-n-1} \right] \left( \hat{d}_2^{\dagger} \right)^n \right. \\
& \left. + \left( \hat{d}_1^{\dagger} \right)^{N-n} \hat{d}_2^{\dagger} \left[ \left( \hat{d}_2^{\dagger} \right)^n \hat{c}_1 + n \left( \hat{d}_2^{\dagger} \right)^{n-1} \right] \right) \ket{0} \\
= & \mu (N-2n) \ket{\psi_{n}^{\perp}} = \lambda_n^{\perp} \ket{\psi_n^{\perp}},
    \end{aligned}
\end{equation}
and thus also of $\hat{L}^{\dagger}$: \begin{equation} 
    \hat{L}^{\dagger} \ket{\psi_n^{\perp}} = \Lambda_n^{\perp} \ket{\psi_n^{\perp}} = \sqrt{\Gamma_c} \left( \lambda_n^{\perp} + \chi \right) \ket{\psi_n^{\perp}}.
\end{equation}
Taking the Hermitian conjugate, this also implies
\begin{equation}
    \bra{\psi_n^{\perp}} \hat{L} = \bra{\psi_n^{\perp}} \Lambda_n^{\perp}.
\end{equation}
Note that for the relevant values of $n = k \pm 1$, we have 
\begin{equation} \label{Lambda_perp}
    \Lambda_{k \pm 1}^{\perp} = \sqrt{\Gamma_c} \left[ \mu (N - 2k \mp 2) + \chi_k \right] = \mp 2 \mu \sqrt{\Gamma_c}.
\end{equation}
Since the eigenvalue $\Lambda_k$ is zero by the construction of $\chi_k$, the denominator of the adiabaticity parameter can be written as
\begin{equation}
    \begin{aligned}
\alpha_{n k} + i \zeta_n &= \frac{\nu + i}{2} \bra{\psi_n^{\perp}} \hat{L}^{\dagger} \hat{L} \ket{\psi_n^{\perp}} \\ 
&= \frac{\nu + i}{2} \bra{\psi_n^{\perp}} \left( \hat{L} \hat{L}^{\dagger} - \left[ \hat{L}, \hat{L}^{\dagger} \right] \right) \ket{\psi_n^{\perp}} \\
&= \frac{\nu + i}{2} \bra{\psi_n^{\perp}} \left[ \left( \Lambda_n^{\perp} \right)^2 +2 \Gamma_c \left( 1 - \mu^4 \right) \hat{J}^z \right] \ket{\psi_n^{\perp}} \\
&= \Gamma_c \left( \nu + i \right) \left[ 2 \mu^2 + \left( 1 - \mu^4 \right) \bra{\psi_n^{\perp}} \hat{J}^z \ket{\psi_n^{\perp}} \right],
    \end{aligned}
\end{equation}
for $n = k \pm 1$, where we have used
\begin{equation}
    \begin{aligned} 
\left[ \hat{L}, \hat{L}^{\dagger} \right] &= \Gamma_c \left[ \left( \tilde{L} + \chi \hat{\mathbb{I}} \right), \left( \tilde{L}^{\dagger} + \chi \hat{\mathbb{I}} \right) \right] = \Gamma_c \left[ \tilde{L}, \tilde{L}^{\dagger} \right] \\
&= \Gamma_c \left[ \left( \hat{J}^- + \mu^2 \hat{J}^+ \right), \left( \hat{J}^+ + \mu^2 \hat{J}^- \right) \right] \\
&= 2 \Gamma_c \left( \mu^4 -1 \right) \hat{J}^z,
    \end{aligned}
\end{equation}
since $\left[ \hat{J}^+, \hat{J}^- \right] = 2 \hat{J}^z$.
We now define
\begin{equation}
    \xi_k = \abs{\frac{4 \bra{\psi_{k \pm 1}^{\perp}} \hat{J}^z \ket{\psi_k}}{\mu \left[ 2 \mu^2 + \left( 1 - \mu^4 \right) \bra{\psi_{k \pm 1}^{\perp}} \hat{J}^z \ket{\psi_{k \pm 1}^{\perp}} \right]}}
\end{equation}
and so the final form of the adiabatic criteria becomes
\begin{equation}
    \Xi_k = \frac{\dot{\mu}}{\Gamma_c \sqrt{1 + \nu^2}} \xi_k \ll 1,
\end{equation}
where we assumed that $\dot{\mu}$ is real and positive.

\subsection{Final Value} \label{FinalValueAdiabaticityParameterCalculation}
We now wish to calculate the final value of the adiabaticity parameter $\Xi_{k,f} \equiv \Xi_k (t_f)$.
Using 
\begin{equation} \label{Jz_on_complementary_state}
    \begin{aligned}
\hat{J}^z \ket{\psi_n^{\perp}} &= - \frac{\left( 2 \mu \right)^N}{2} \mathcal{N}_n^{\perp} \left[ \hat{d}_2^{\dagger} \hat{c}_1 + \hat{d}_1^{\dagger} \hat{c}_2 \right] \left( \hat{d}_1^{\dagger} \right)^{N-n} \left( \hat{d}_2^{\dagger} \right)^n \ket{0} \\
&= - \frac{\mathcal{N}_n^{\perp}}{2} \left[ \frac{N-n}{\mathcal{N}_{n+1}^{\perp}} \ket{\psi_{n+1}^{\perp}} + \frac{n}{\mathcal{N}_{n-1}^{\perp}} \ket{\psi_{n-1}^{\perp}} \right],
    \end{aligned}
\end{equation}
we first find
\begin{equation} \label{AdiabaticityParameterDenominator}
    \begin{aligned} 
\alpha_{nk} + i \zeta_n = & \Gamma_c \left( \nu + i \right) \left[ 2 \mu^2 + \frac{\mathcal{N}_n^{\perp}}{2} \left( \mu^4 - 1 \right) \times \right. \\
& \left. \left( \frac{N-n}{\mathcal{N}_{n+1}^{\perp}} \ip{\psi_n^{\perp}}{\psi_{n+1}^{\perp}} + \frac{n}{\mathcal{N}_{n-1}^{\perp}} \ip{\psi_n^{\perp}}{\psi_{n-1}^{\perp}} \right) \right].
    \end{aligned}
\end{equation}
In the case $\mu = 1$, the overlap sum Eq.~\eqref{EigenstateOverlaps} reduces to
\begin{equation}
    \ip{\psi_{k'}}{\psi_k} = 2^N \abs{\mathcal{N}_k}^2 \left( N - k \right)! k! \delta_{k, k'} = \delta_{k, k'},
\end{equation}
as $a_{\perp} = 1$ and $a_1 = 0$ picks out the $i = 0$ term in the sum and sets $\ip{\psi_{k'}}{\psi_k} = 0$ when $k \neq k'$, and similarly for the complementary states.
This combined with Eq.~\eqref{AdiabaticityParameterDenominator} allows us to obtain
\begin{equation}
    \alpha_{n k} + i \zeta_n = 2 \Gamma_c \left( \nu + i \right),
\end{equation}
and thus
\begin{equation}
    \begin{aligned} 
\Xi_{k,f} = & \max_{n = k \pm 1} \left[ \frac{4 \dot{\mu} 2^{N-1}}{2 \Gamma_c \sqrt{1 + \nu^2} \sqrt{2^N \left( N - k \right)! k!}} \times \right. \\
& \left. \left( \frac{\left( N - k \right)! \left( k + 1 \right)!}{\sqrt{2^N \left( N - k - 1 \right)! \left( k + 1 \right)!}} \delta_{n, k+1} \right. \right. \\
& \left. \left. + \frac{\left( N - k + 1 \right)! k!}{\sqrt{2^N \left( N - k + 1 \right)! \left( k - 1 \right)!}} \delta_{n, k-1} \right) \right],
    \end{aligned}
\end{equation}
where we have assumed that $\mu (t_f) = 1$.
Simplifying, we find
\begin{equation}
    \begin{aligned}
\Xi_{k,f} = & \max_{k \pm 1} \left[ \frac{\dot{\mu}}{\Gamma_c \sqrt{1 + \nu^2}} \left( \sqrt{(N-k)(k+1)} \delta_{n, k+1} \right. \right. \\
& \left. \left. + \sqrt{(N-k+1)k} \delta_{n, k-1} \right) \right],
    \end{aligned}
\end{equation}
which we rewrite as 
\begin{equation}
    \Xi_{k,f} = \frac{\dot{\mu}}{\Gamma_c \sqrt{1 + \nu^2}} \xi_{k,f},
\end{equation}
where we have defined
\begin{equation}
\xi_{k,f} = 
\begin{cases}
    & \sqrt{(N-k)(k+1)}, \quad k < \frac{N}{2} \\
    & \sqrt{(N-k+1)k}, \quad k \geq \frac{N}{2}
\end{cases}.
\end{equation}

\subsection{Single Particle Value}
In the case $N=1$, we have
\begin{equation}
    \ket{\psi_0} = \mathcal{N}_0 \left( \mu \hat{b}_{\uparrow}^{\dagger} + \hat{b}_{\downarrow}^{\dagger} \right) \ket{0}, \quad \ket{\psi_1^{\perp}} = \mathcal{N}_1^{\perp} \left( -\hat{b}_{\uparrow}^{\dagger} + \mu \hat{b}_{\downarrow}^{\dagger} \right) \ket{0},
\end{equation}
with $\mathcal{N}_0 = \mathcal{N}_1^{\perp} = 1/\sqrt{1 + \mu^2}$.
Using Eq.~\eqref{Jz_on_eigenstate}, we project the complimentary state on the right to find
\begin{equation}
    \bra{\psi_n^{\perp}} \hat{J}^z \ket{\psi_k} = - \frac{\mu}{1 + \mu^2},
\end{equation}
as well as using Eq.~\eqref{Jz_on_complementary_state} to write 
\begin{equation}
    \bra{\psi_1^{\perp}} \hat{J}^z \ket{\psi_1^{\perp}} = - \frac{\mathcal{N}_1^{\perp}}{2 \mathcal{N}_0^{\perp}} \ip{\psi_1^{\perp}}{\psi_0^{\perp}} = - \frac{1}{2} \frac{\mu^2 - 1}{\mu^2 + 1},
\end{equation}
where we have used $\mathcal{N}_0^{\perp} = 1/\sqrt{1 + \mu^2}$ and
\begin{equation}
    \begin{aligned}
\ip{\psi_1^{\perp}}{\psi_0^{\perp}} = (2 \mu)^2 \mathcal{N}_1^{\perp} \mathcal{N}_0^{\perp}\bra{0} \hat{d}_2 \hat{d}_1^{\dagger} \ket{0} = \mathcal{N}_1^{\perp} \mathcal{N}_0^{\perp} = \frac{\mu^2 - 1}{\mu^2 + 1}.
    \end{aligned}
\end{equation}
We therefore, from Eqs.~\eqref{xi_k} and~\eqref{Xi_k}, obtain
\begin{equation}
    \begin{aligned}
\Xi_0 &= \abs{- \frac{4 \dot{\mu} \mu}{1 + \mu^2} \frac{1}{\mu \Gamma_c \left( \nu + i \right) \left[ 2 \mu^2 - (1 - \mu^4) \frac{1}{2} \frac{\mu^2 - 1}{\mu^2 + 1} \right]}} \\
&= \abs{\frac{- 4 \dot{\mu}}{\Gamma_c \left( \nu + i \right) \left[ 2 \mu^2 \left( 1 + \mu^2 \right) - \frac{\left( 1 - \mu^4 \right) \left( \mu^2 - 1 \right)}{2} \right]}}.
\end{aligned}
\end{equation}

\end{document}